\documentclass[structabstract]{aa}  

\usepackage{natbib}
\bibpunct{(}{)}{;}{a}{}{,} % to follow the A&A style
\usepackage{graphicx}
\usepackage{txfonts}
\begin{document}
   \title{On the importance of background subtraction in the analysis of coronal loops observed with TRACE}

   \subtitle{}

   \author{S.Terzo
          \inst{1}
          \and
          F.Reale\inst{1,2}%\fnmsep%\thanks{Just to show the usage
          %of the elements in the author field}
          }

   \institute{Dipartimento di Scienze Fisiche ed Astronomiche, Universit\`a
	             degli studi di Palermo,
              Via Archirafi 36, 90123, Palermo, Italy\\
              \email{terzo@astropa.unipa.it}
         \and
             INAF Osservatorio Astronomico di Palermo,
	     Piazza del Parlamento 1, 90134 Palermo, Italy\\
             \email{reale@astropa.unipa.it}
             %\thanks{The university of heaven temporarily does not
              %       accept e-mails}
             }

   \date{Received 15/10/2009 ; accepted 05/02/2010}

% \abstract{}{}{}{}{} 
% 5 {} token are mandatory
 
  \abstract
{}
{ In the framework of TRACE coronal observations, we compare the analysis and diagnostics of a loop after subtracting the background with two different and independent methods.} %leave it empty if necessary 
{The dataset includes sequences of images in the 171~{\AA}~, 195~{\AA}~ filter bands of TRACE. One background subtraction method consists in taking as background values those obtained from interpolation between concentric strips around the analyzed loop. The other method is a pixel-to-pixel subtraction of the final image when the loop had completely faded out, already used by \citealp{Reale_2006}.} 
{ We compare the emission distributions along the loop obtained with the two methods and find that they are considerably different. We find differences as well in the related derive filter ratio and temperature profiles. In particular, the pixel-to-pixel subtraction leads to coherent diagnostics of a cooling loop. With the other subtraction the diagnostics are much less clear.} 
{ The background subtraction is a delicate issue in the analysis of a loop. The pixel-to-pixel subtraction appears to be more reliable, but its application is not always possible. Subtraction from interpolation between surrounding regions can produce higher systematic errors, because of intersecting structures and of the large amount of subtracted emission in TRACE observations.  
}

   \maketitle
%
%________________________________________________________________

\begin{figure}[]       %%%%%%%%%%%%%%%%%% FIGURE 1
	\centering	
	\includegraphics[width=8.2cm]{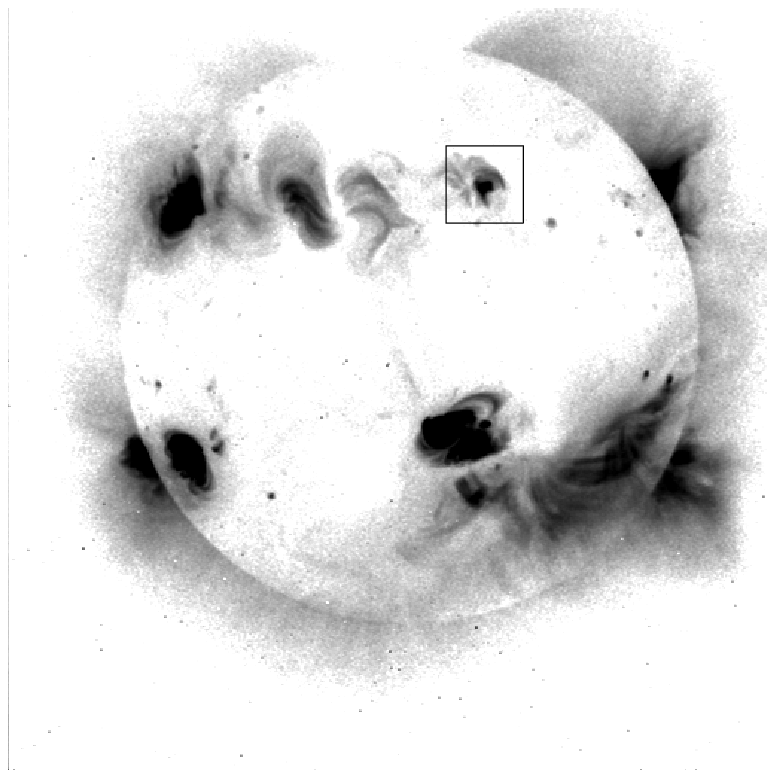}
	\includegraphics[width=9.0 cm]{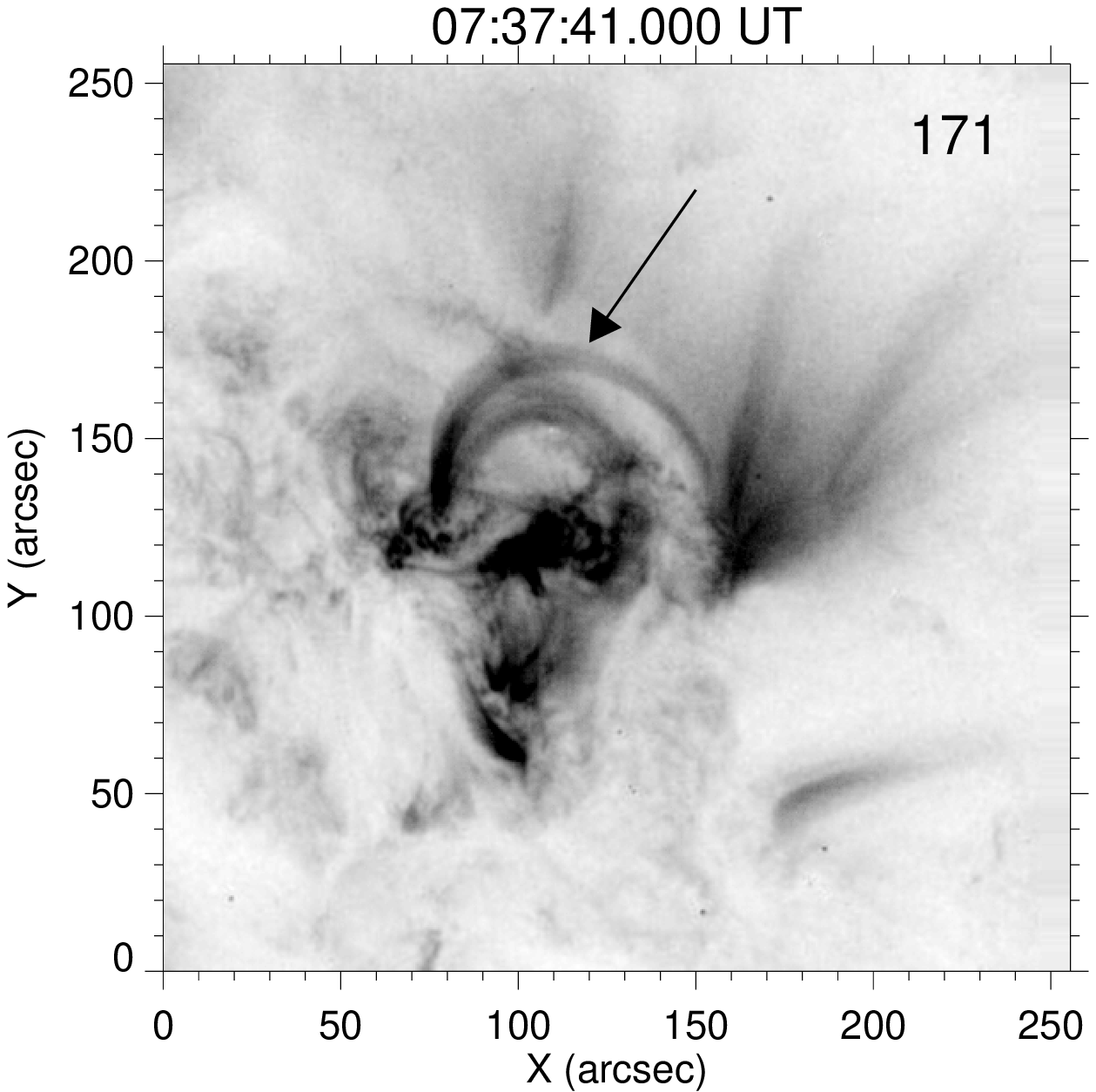}

\caption{Loop region as observed with TRACE ($256\times256\;pixels$ image) in the 171~\AA~ filter at the labelled time ({\it lower panel}). The loop under analysis is pointed with an arrow. The loop region is located in the inset of the Yohkoh/SXT full disk image ({\it upper panel}). The grey scale is inverted and linear for the TRACE image ($\leq 8$ DN/s/pix) and inverted and logarithmic for the Yohkoh image (between 10 and 150 DN/s/pix).}
	\label{immagini_dati}
\end{figure}

\section{Introduction}

An important issue in the data analysis of coronal loops observations is background subtraction. Recent results ( \citealp{DelZanna_2003}, \citealp{Testa_2002}, \citealp{Schmelz_2003}, \citealp{Aschw_2005}, \citealp{Reale_2006}, \citealp{Aschw_2008ApJ}) have established the importance of separating the actual loop plasma from the diffuse foreground and background emission, that results from unresolved coronal structures and instrumental effects. The need of background subtraction arises from the presence of many overlapping bright structures and of diffuse emission, nearby or along the line of sight, as well as from stray light (\citealp{DeForest_2009ApJ}). The accurate extraction of the emission along the loop is necessary to apply standard diagnostic methods (like filters ratio) to derive physical quantities, such as temperature, or even to apply more detailed loop models. Wherever the background is non-negligible compared with the loop intensity, it would seriously affect the extracted intensity along the loop. There is no standard and generally accepted method of background subtraction; the procedure is then ``operator-sensitive'' and so are in turn the results. 

The issue is critical especially in the analysis of observations where the background is a significant fraction of the signal. This happens, for instance, in observation made with \emph{TRACE} (e.g.\citealp{Schmelz_2003}, \citealp{Aschw_2005}, \citealp{Reale_2006}). Here we explore the dependence and sensitivity of the results on the background subtraction method, by comparing two methods applied to the same loop, observed with \emph{TRACE} in more than one filter band. We take advantage of the unique opportunity to apply two different and independent methods of background subtraction on the same dataset. One of the methods (\citealp{Reale_2006} - hereafter RC06) has an objective support from the facts that we use as background an image of the loop region when the loop is absent, and that background variations are estimated to be small throughout the observation.  

In section 2 we describe the data analysis and background subtraction. Section 3 shows and compares the results with the two different background subtractions, including implications on temperature diagnostics with filter ratios. In section 4 we discuss the results.

\section{Data analysis}

The {\it Transition Region and Coronal Explorer} (TRACE) is a NASA Small Explorer (SMEX) mission to image the solar corona and transition region at high angular and temporal resolution, operating since 1998 \citep{TRACE_1999}. We analyze a \emph{TRACE} observation of May 13 1998, with a 3.5h time sequence of $1024\times1024$ pixels full resolution image in two of TRACE filters (171~{\AA}~, 195~{\AA}). This is the same data set selected and analyzed by RC06. The same $512\times512$ pixels region of the whole field of view has been extracted for the analysis. From these data four images were selected, in each filter, at the following times: 06:36:57 UT, 06:59:35 UT, 07:39:26 UT, 08:29:34 UT for 171~{\AA} filter, and 06:37:18 UT, 07:00:06 UT, 07:39:47 UT, 08:30:06 UT for 195~{\AA} filter. The filters have different sensitivity to temperature of observed plasma. Observations taken at the same time with different filters provide information on the plasma temperature. 
In particular, the ratio of the emission in two different filter bands, compared with the ratio of the response functions of the filters, gives the temperature of the emitting plasma. Since the filter ratio may be altered by the diffuse emission, by the signals from other structures co-aligned along the line of sight and by the stray light, it is very important to estimate and remove such background emission. In TRACE observations the diffuse emission is typically very high and might represent most of the signal.

\begin{figure*}[]       %%%%%%%%%%%%%%%%%% FIGURE 2
	\centering	
	\includegraphics[width=4.4cm]{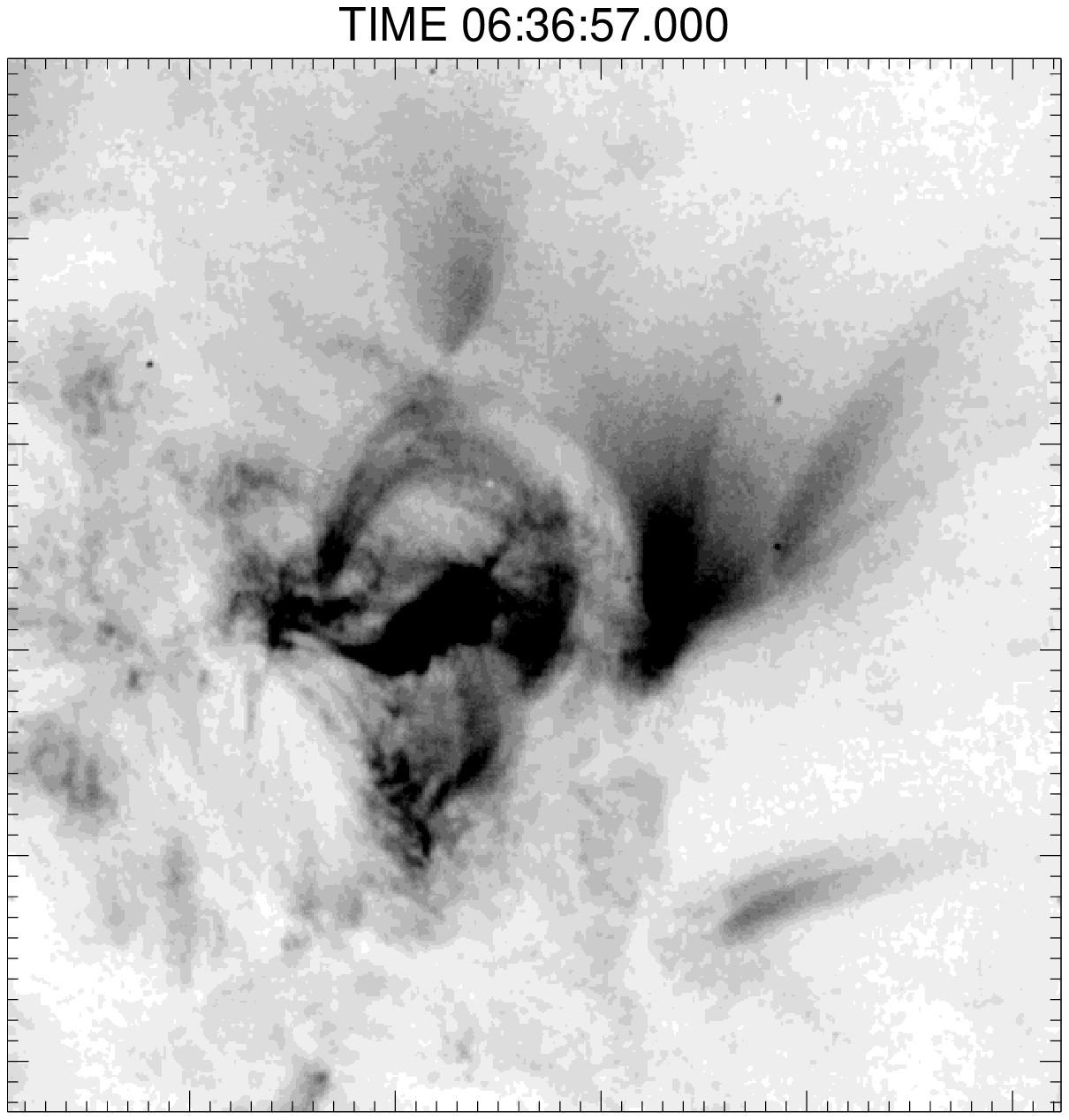}
	\includegraphics[width=4.4cm]{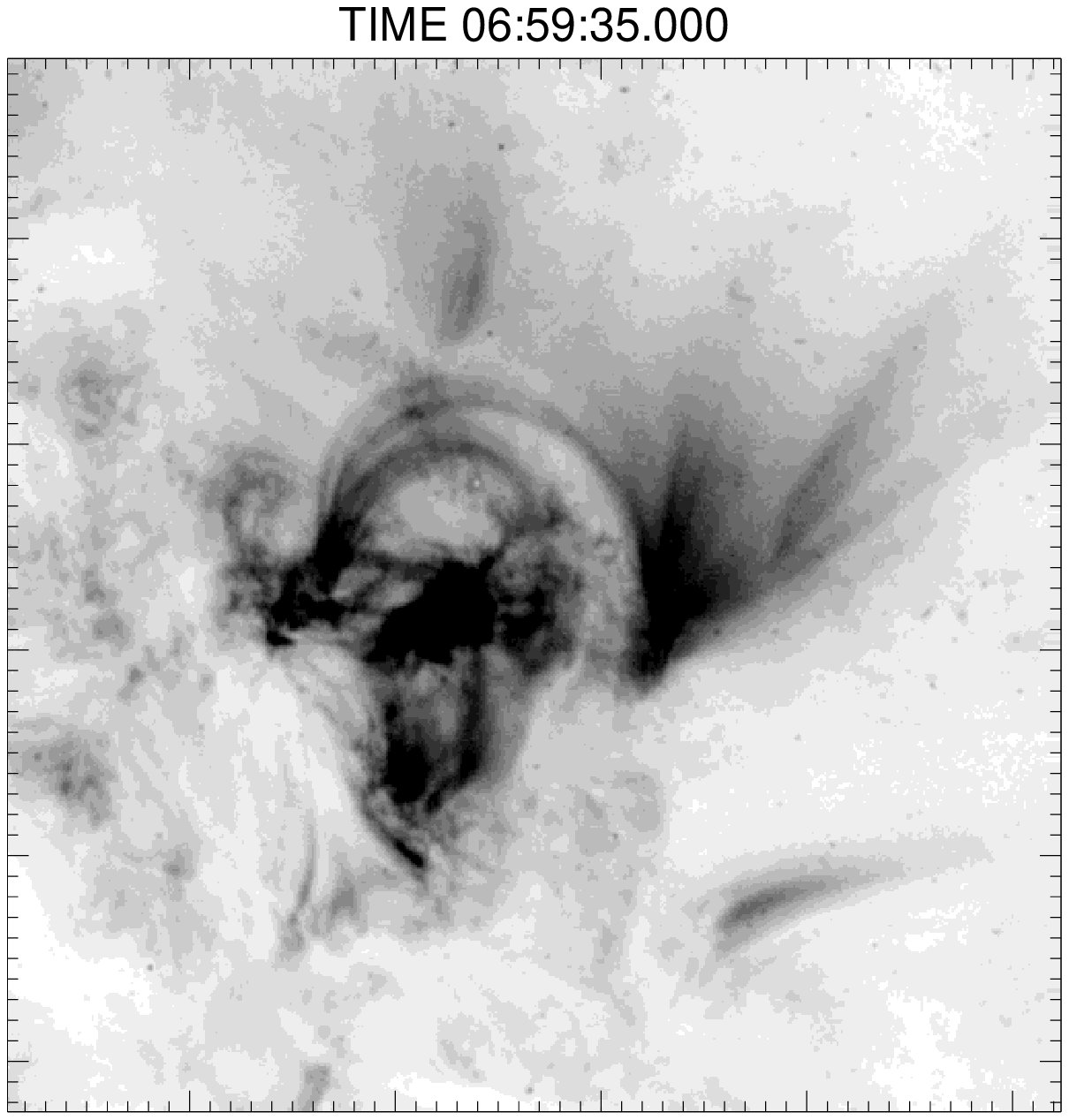}
	\includegraphics[width=4.4cm]{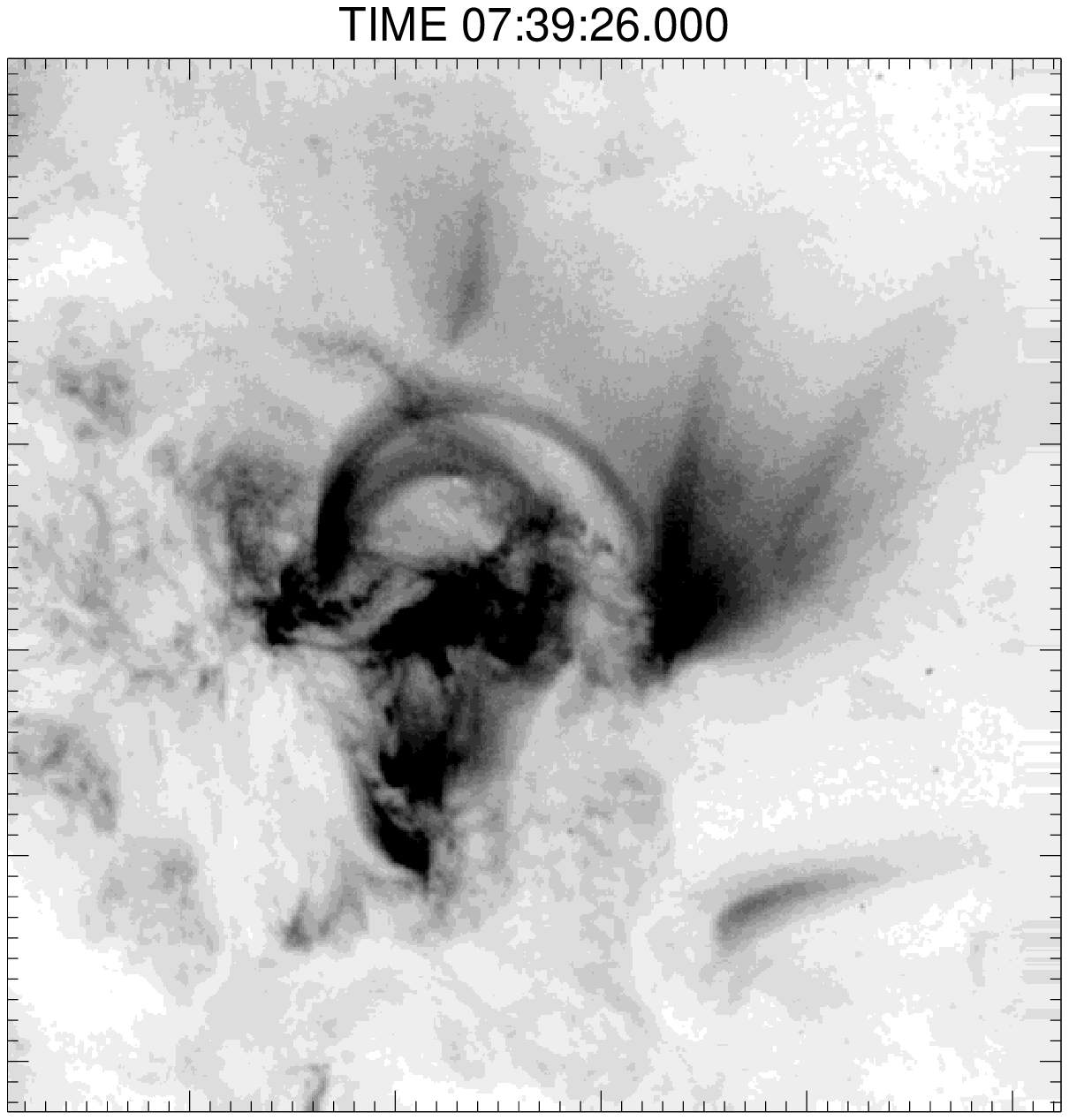}
	\includegraphics[width=4.4cm]{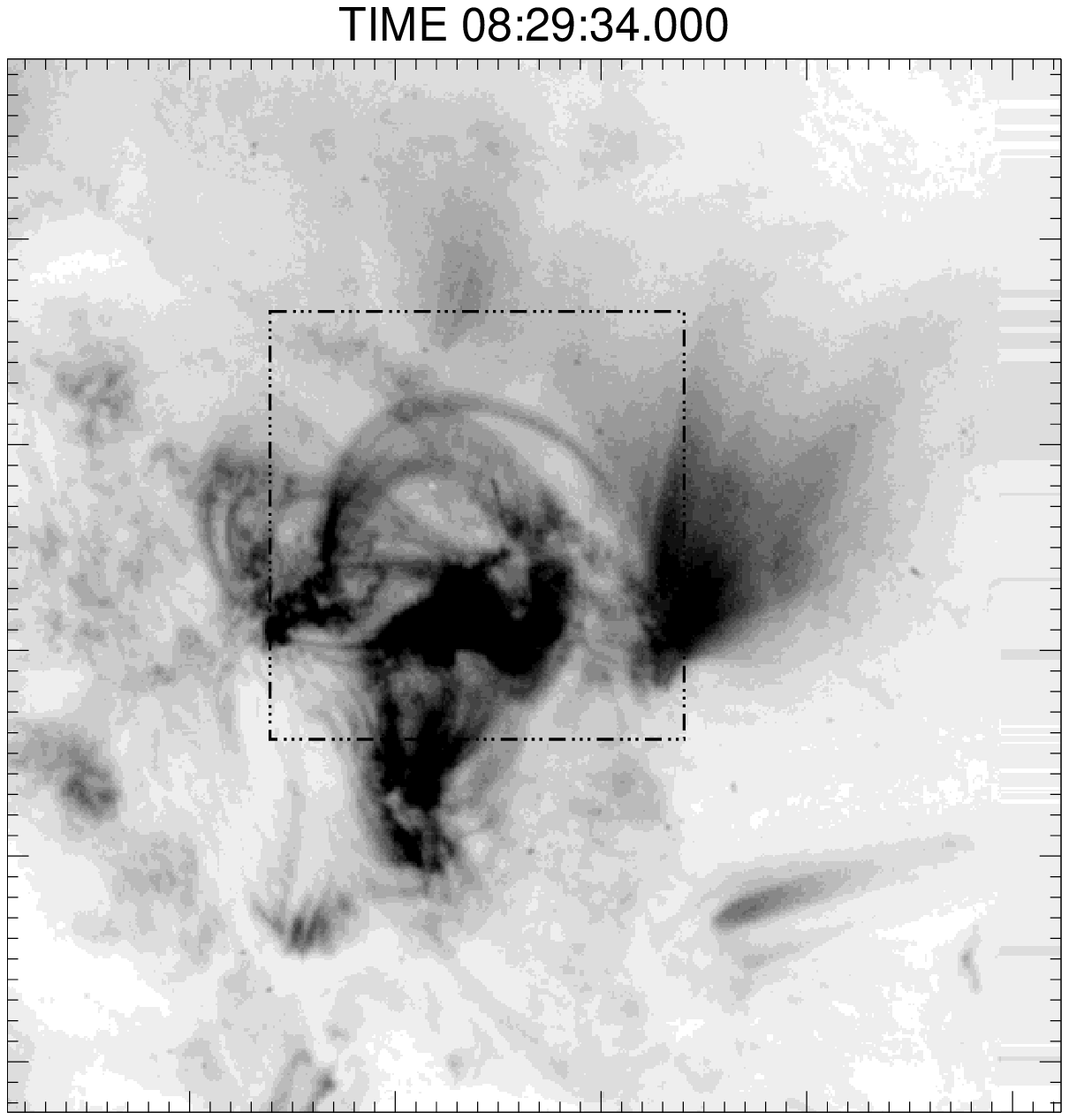} \\

	\includegraphics[width=4.4cm]{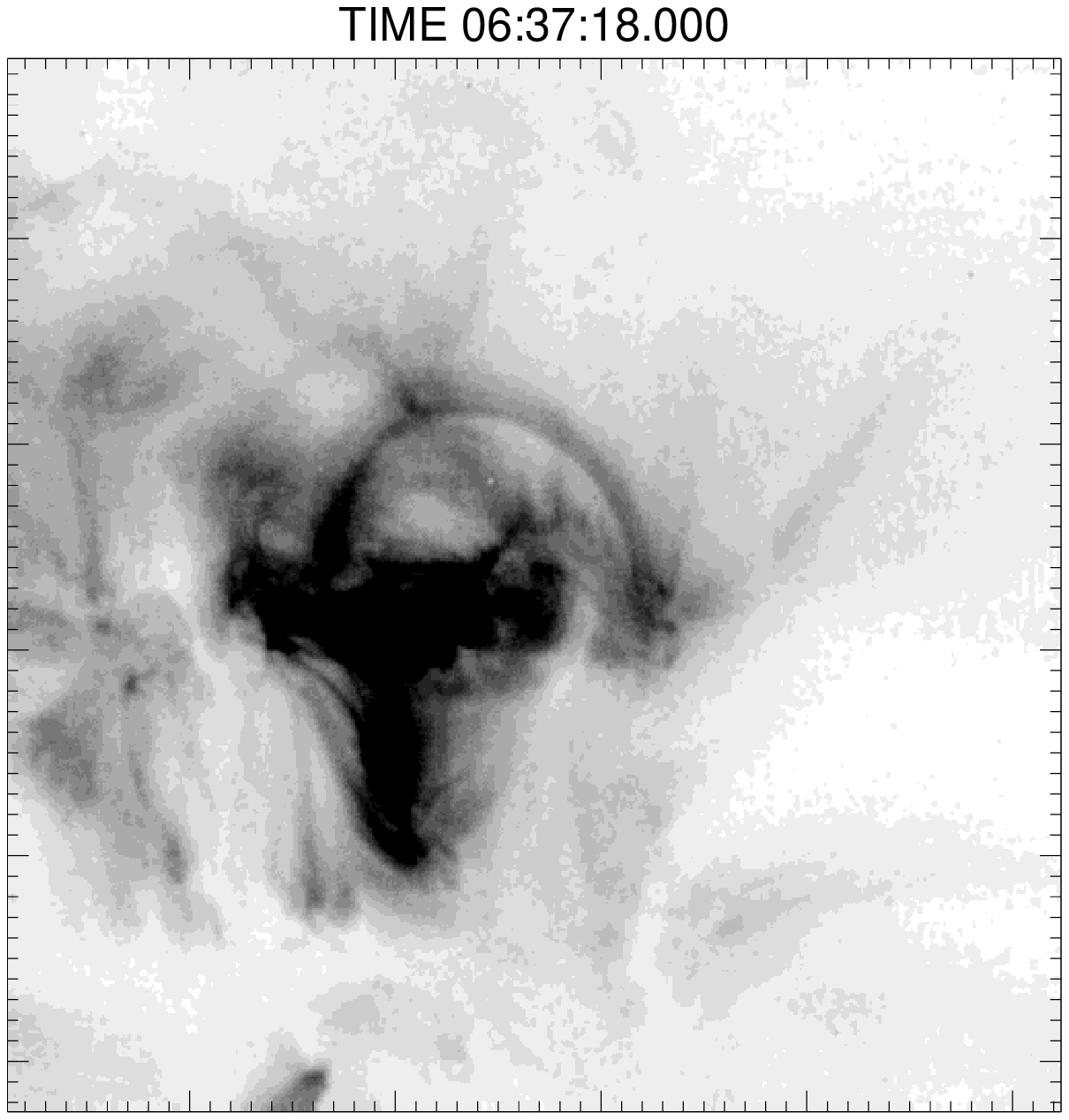}
	\includegraphics[width=4.4cm]{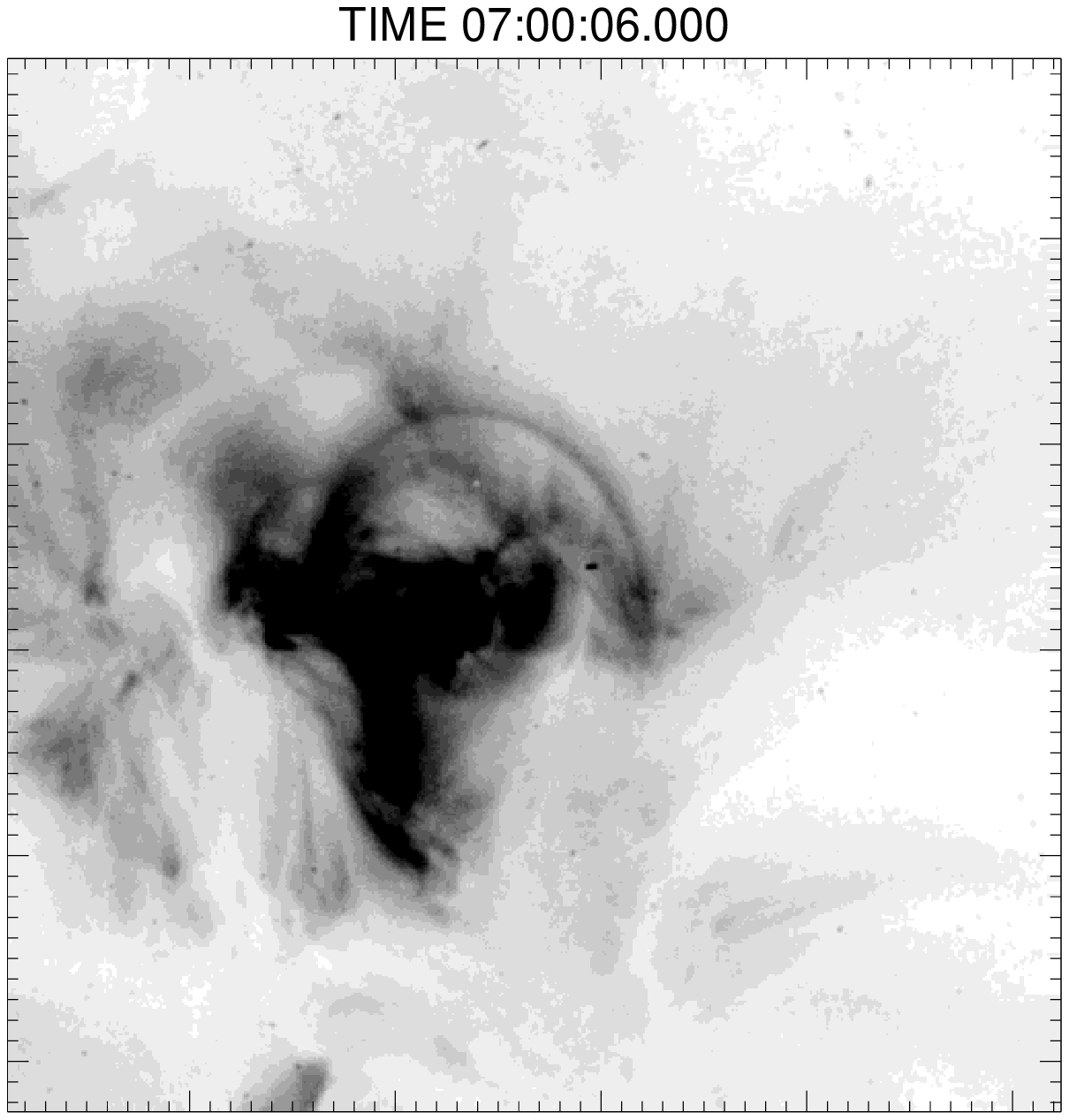}
	\includegraphics[width=4.4cm]{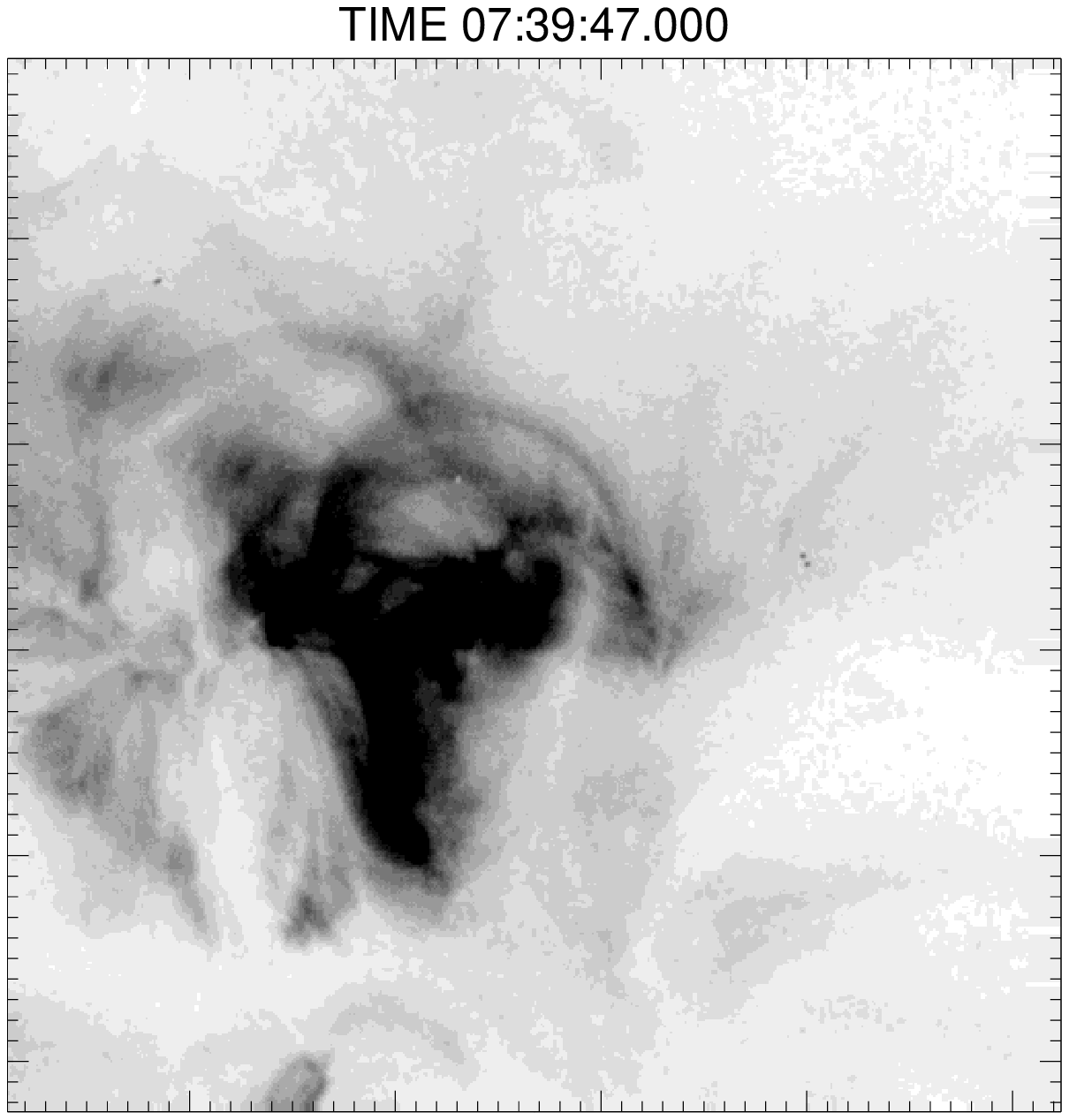}
	\includegraphics[width=4.4cm]{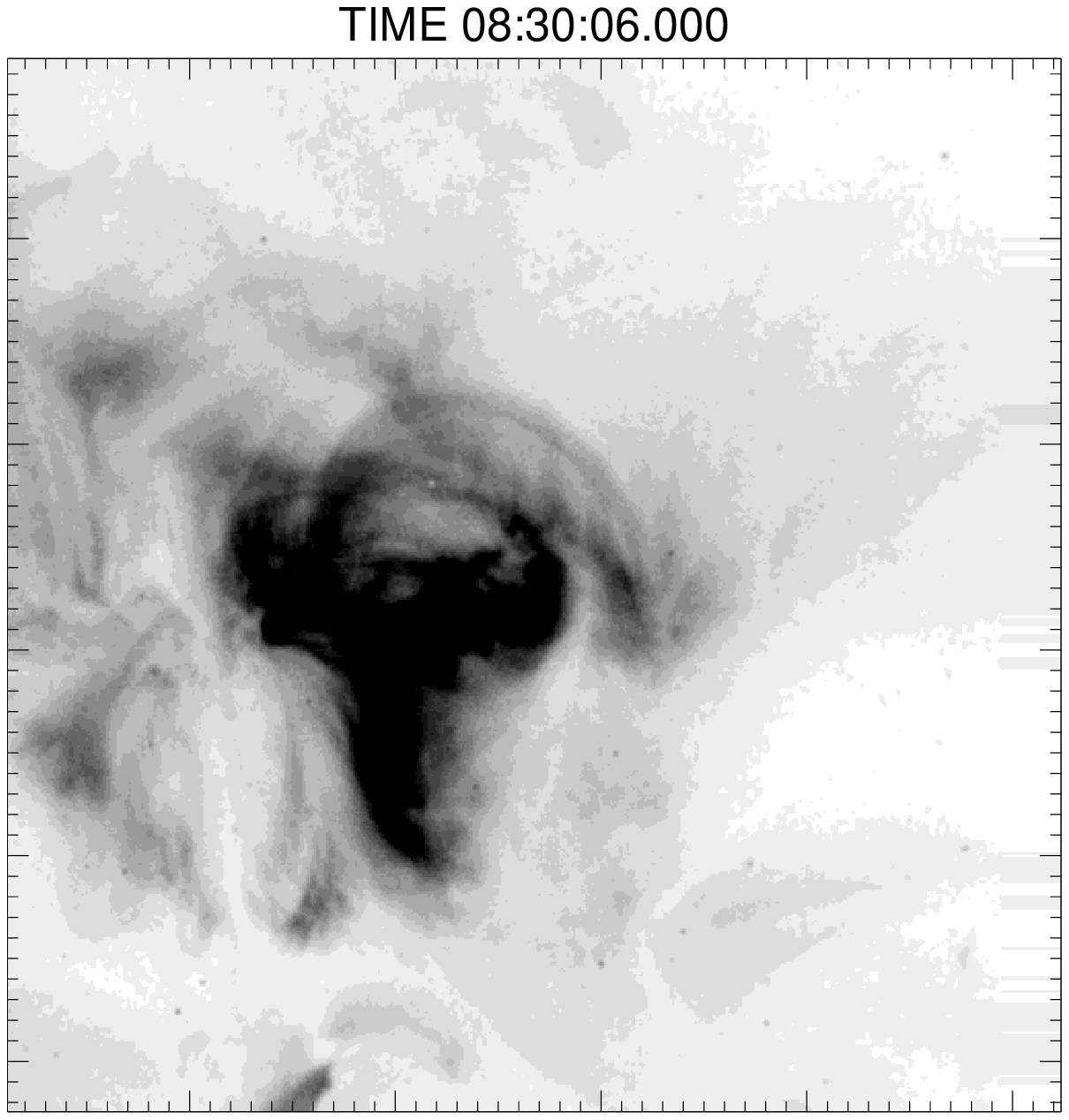}
\centering
	
\caption{ Loop region {\bf (top row: 171~{\AA}~; bottom row: 195~{\AA}~)} at subsequent times {\bf (from left to right)}: 06:36:57.000 UT, 06:59:35.000 UT, 07:39:26.000 UT, 08:29:34.000 UT for 171~{\AA}~ and  06:37:18.000 UT, 07:00:06.000 UT, 07:39:47.000 UT, 08:30:06.000 UT for 195~{\AA}. The grey scale is inverted and linear between 0.4 and 7 DN s$^{-1}$ pix$^{-1}$ for all images. The frame in the top right image is the part shown in Fig.~\ref{ratio-panels}}
	\label{F-4panels-1}
\end{figure*}

\subsection{Loop analysis}

\begin{figure*}[]       %%%%%%%%%%%%%%%%%% FIGURE 3
	\centering
	\includegraphics[width=4.4cm]{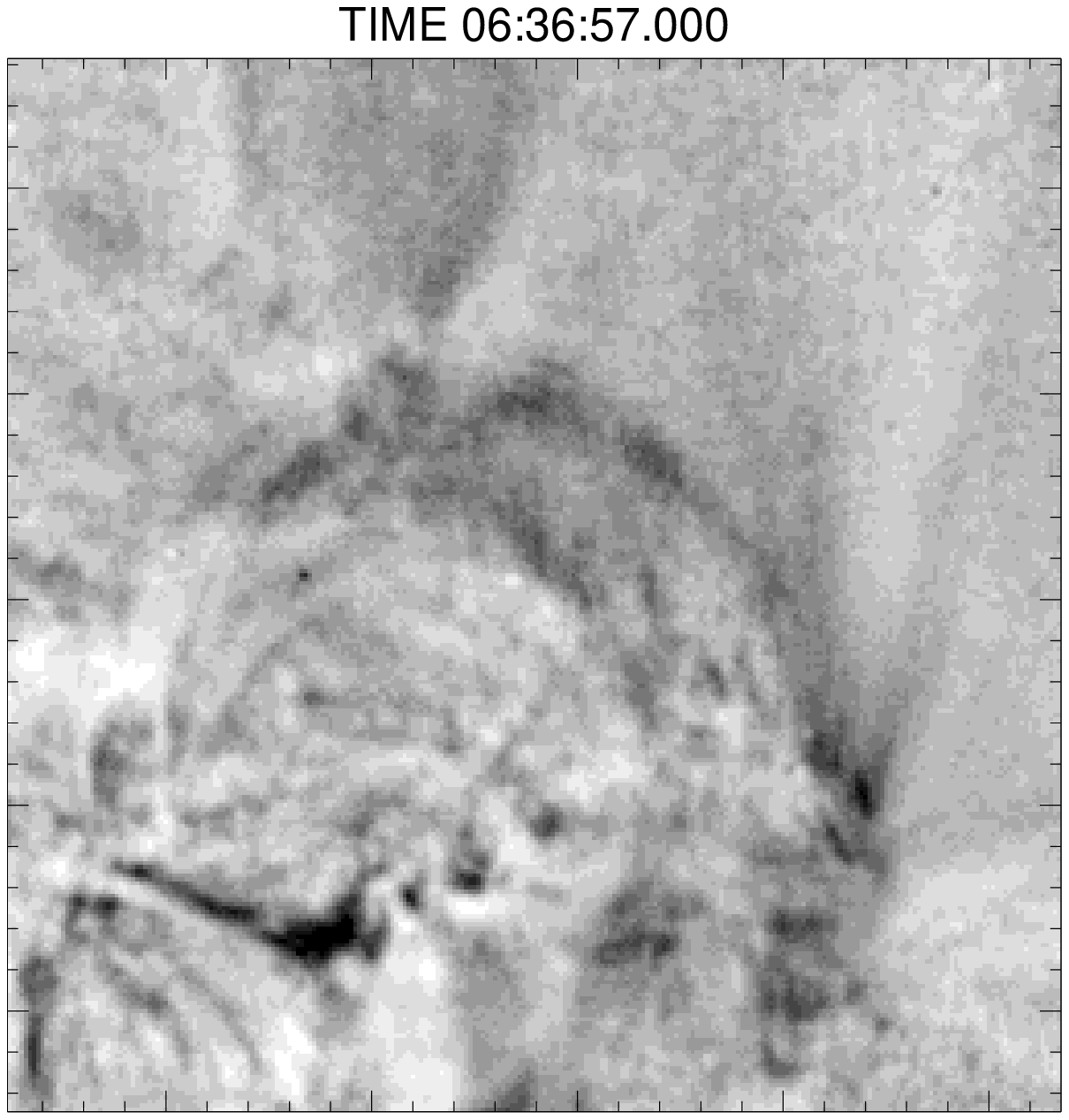}
	\includegraphics[width=4.4cm]{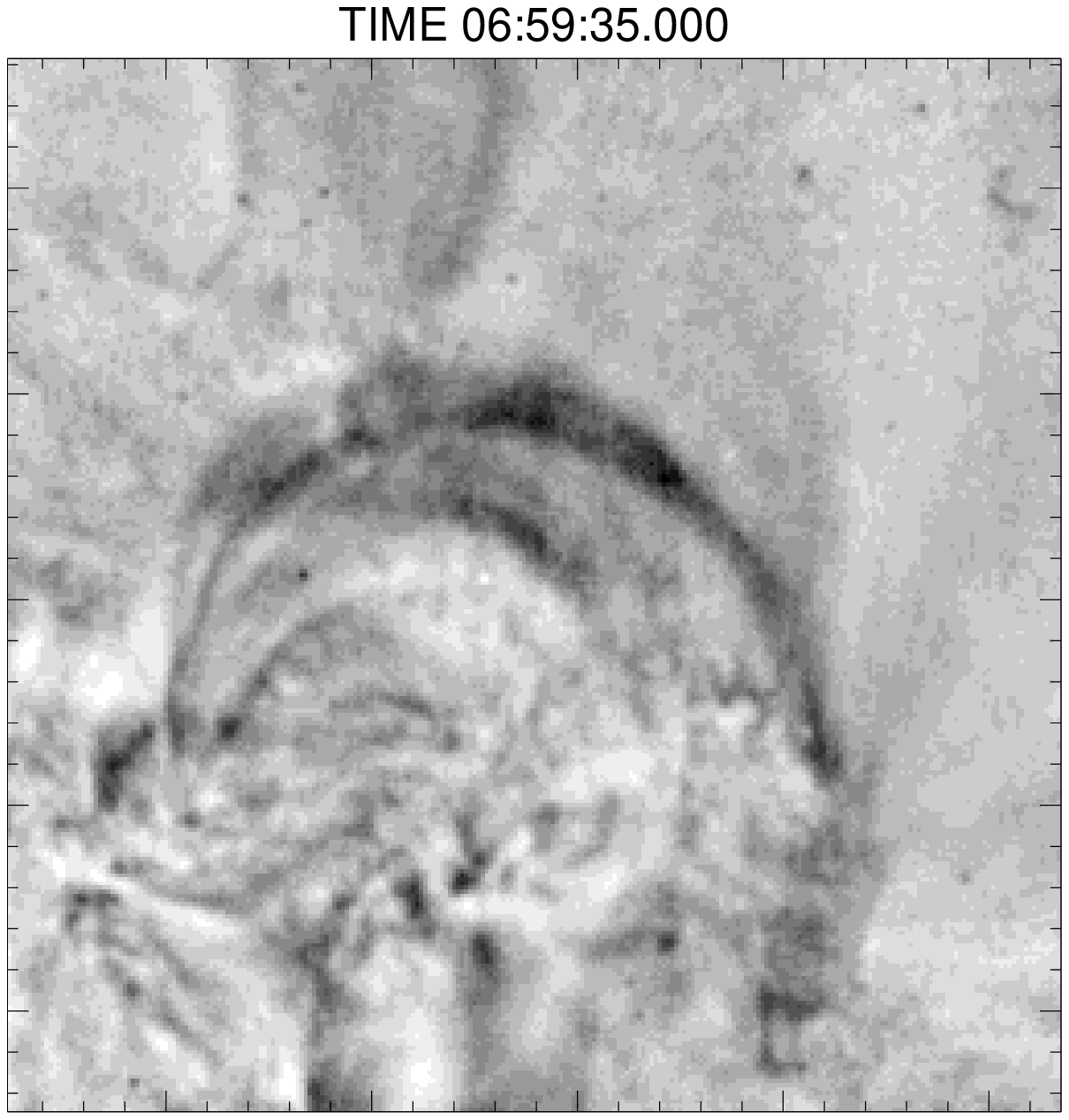}
	\includegraphics[width=4.4cm]{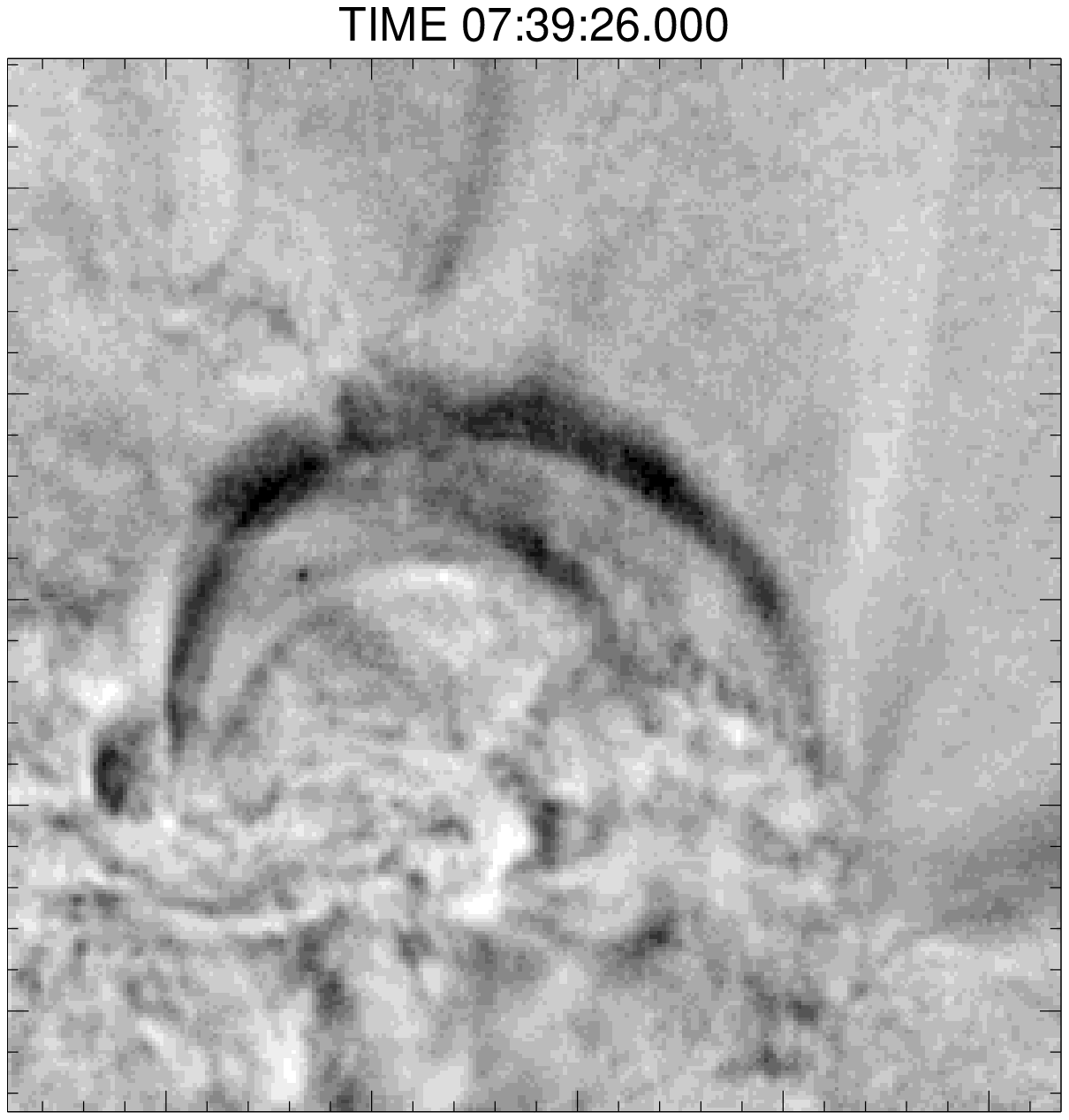}
	\includegraphics[width=4.4cm]{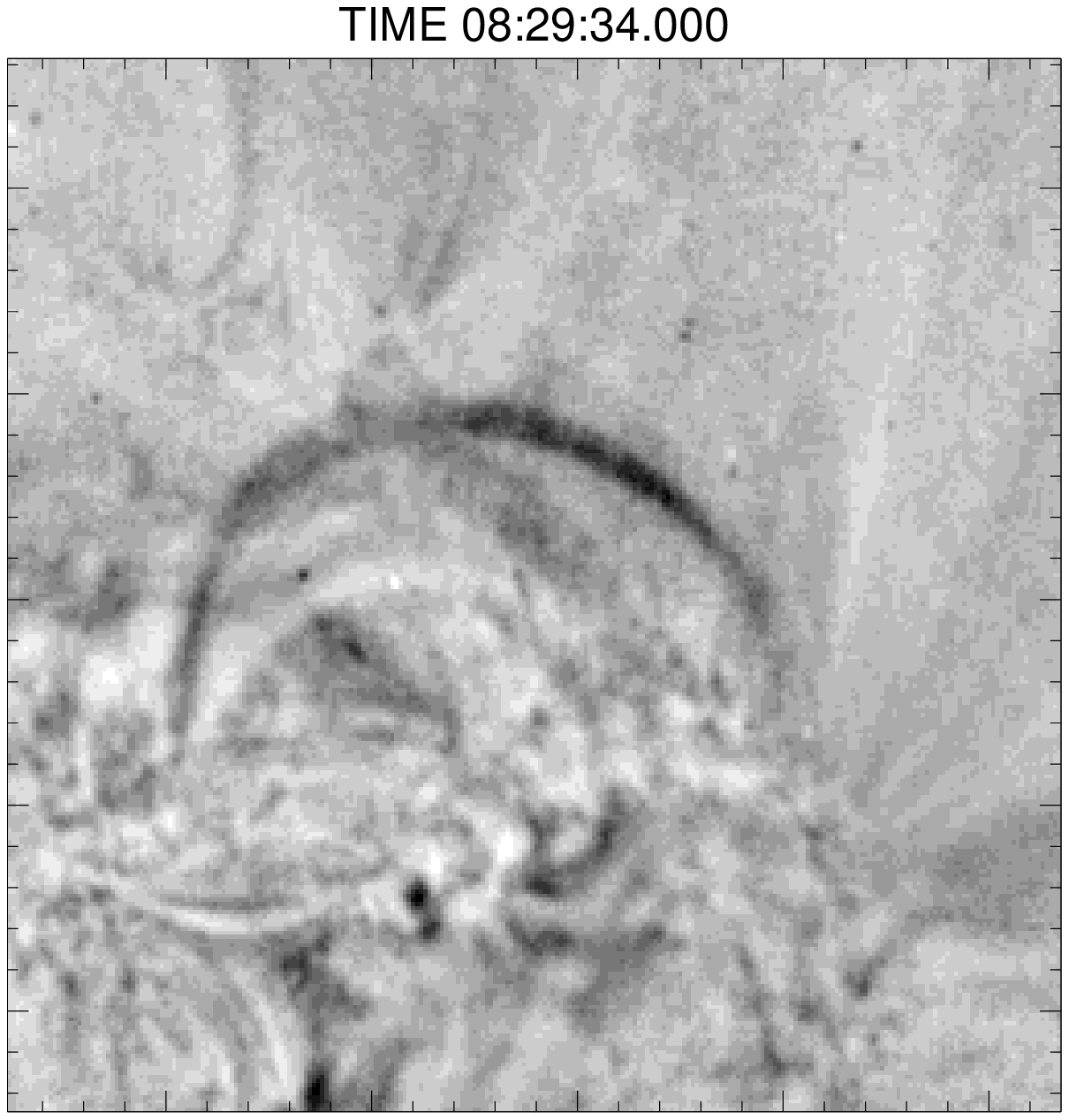} \\

	\includegraphics[width=4.4cm]{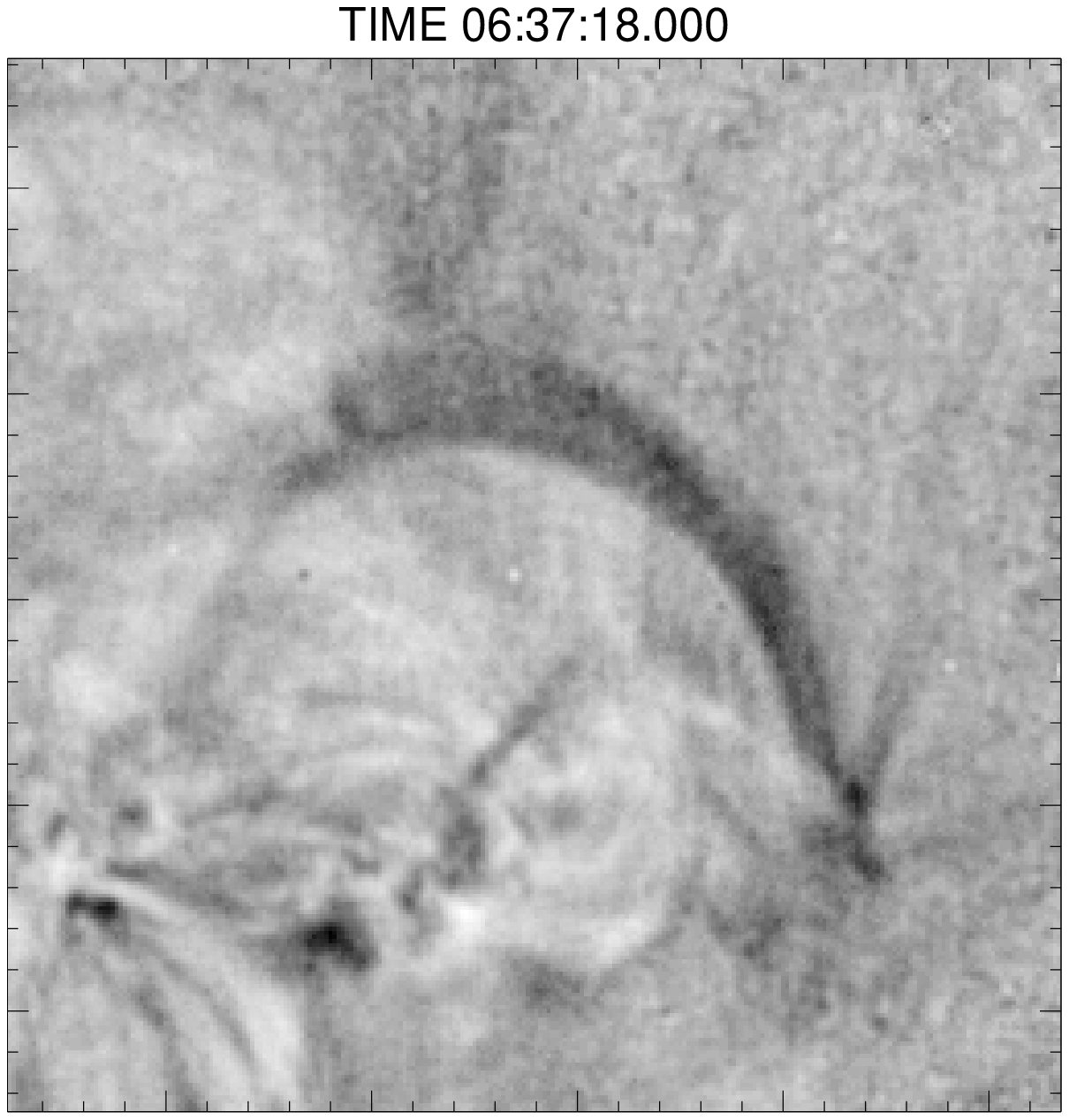}
	\includegraphics[width=4.4cm]{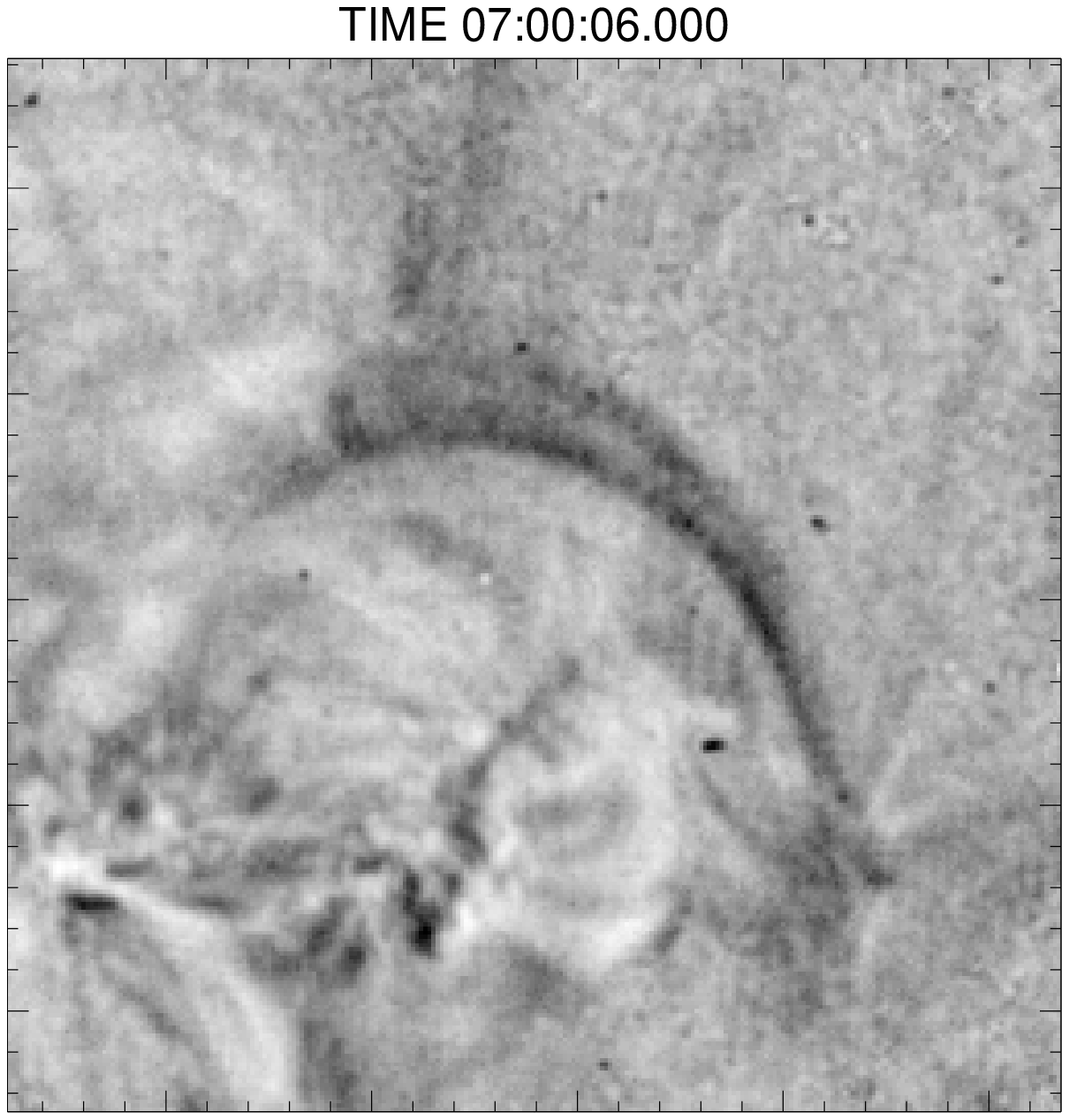}
	\includegraphics[width=4.4cm]{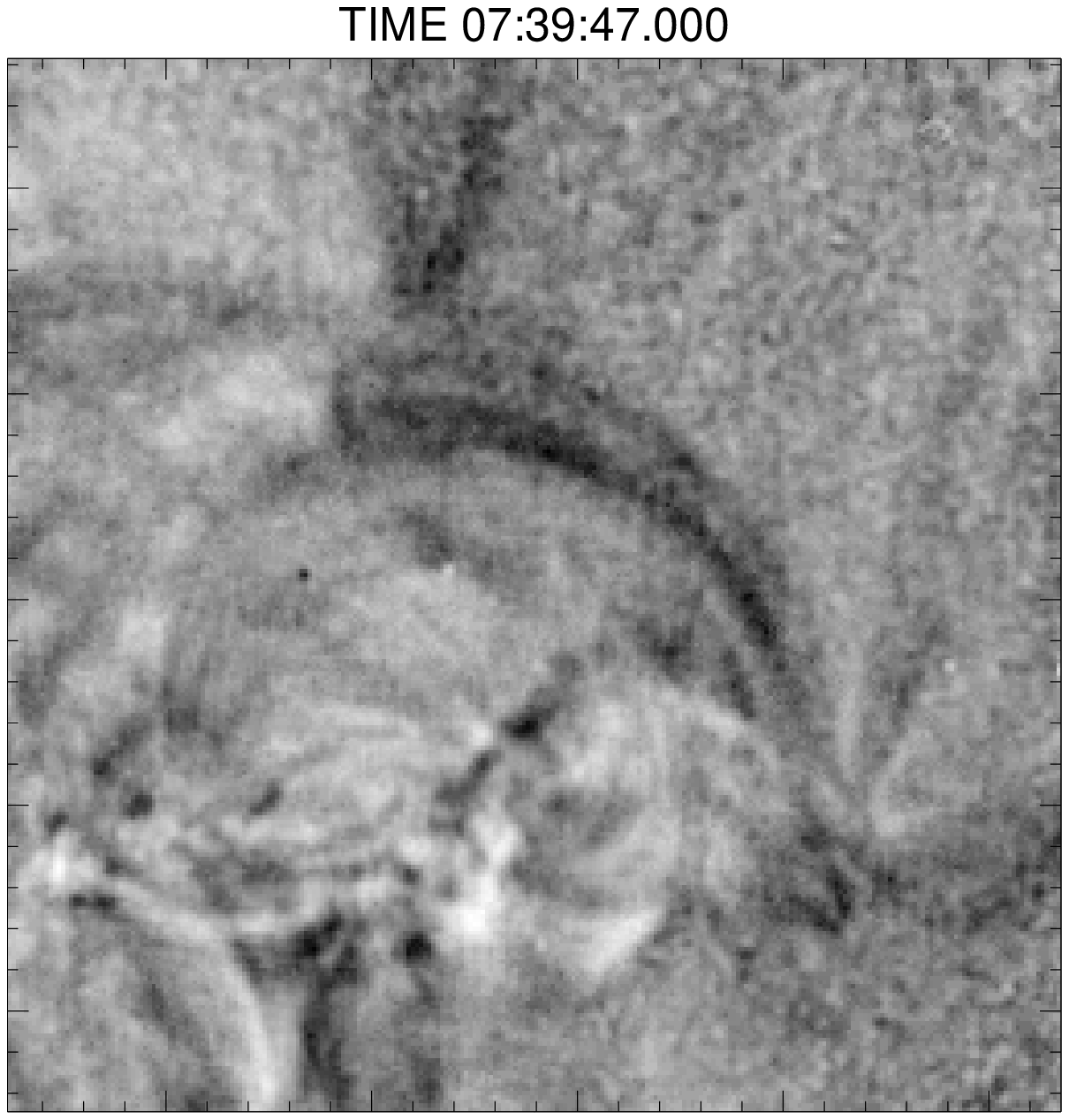}
	\includegraphics[width=4.4cm]{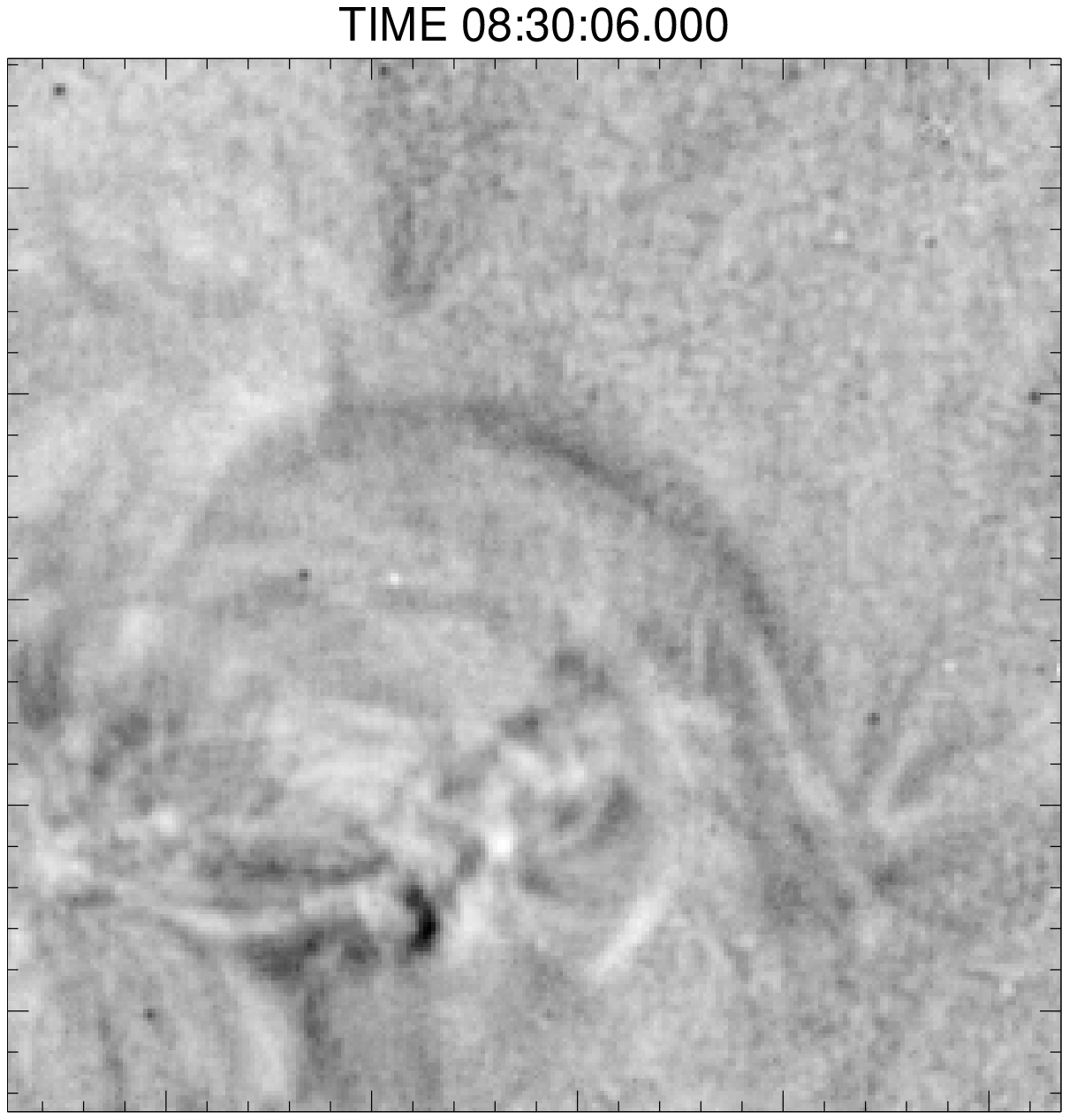}	

\caption{ Zooms of the loop region in Fig.~\ref{F-4panels-1} (frame in the top right image of Fig.~\ref{F-4panels-1}). The images (top row: 171~{\AA}~; bottom row: 195~{\AA}~) are normalized to the final image of the sequence (at 10:00 UT) where the loop is no longer visible, to enhance the loop evolution. The grey scale is linear between 0.4 and 2.5 (1 means that the emission value is the same as that of the final image).}
	\label{ratio-panels}
\end{figure*}

We analyze the same loop as in RC06. The loop is selected on the TRACE images (Fig.\ref{immagini_dati}), and appears as an entire loop in several 171~{\AA}~ filter images  (Fig.\ref{F-4panels-1}). It is clearly visible as well in the 195~{\AA}~ filter band  (Fig.\ref{F-4panels-1}). The loop is bright, i.e. observed with good count statistics and with a high contrast over the background. It was selected as far as possible free from other structures intersecting along the line of sight. The loop evolves in both filterbands. In the 171~{\AA}~ band it is faint initially, then it brightens more reaching a peak of intensity around 07:30 UT and then it fades again, and disappears at the end of the sequence ($\sim$ 10:00 UT, RC06). In the 195~{\AA}~ band the loop is brightest initially and progressively fades out. This loop evolution is even more clear in Fig.~\ref{ratio-panels}, which shows images of the loop region normalized to the final image at 10 UT, where the loop is not longer visible. Here we revisit the data at the four selected times. The data were treated with the standard procedures for \emph{TRACE} data processing contained in the {\it Solar SoftWare} ({\it ssw}) and the images were co-aligned with a standard cross-correlation technique.

\subsection{Background subtraction}

\begin{figure}[!h]       %%%%%%%%%%%%%%%%%% FIGURE 4
	\centering	
	\includegraphics[width=9.8cm]{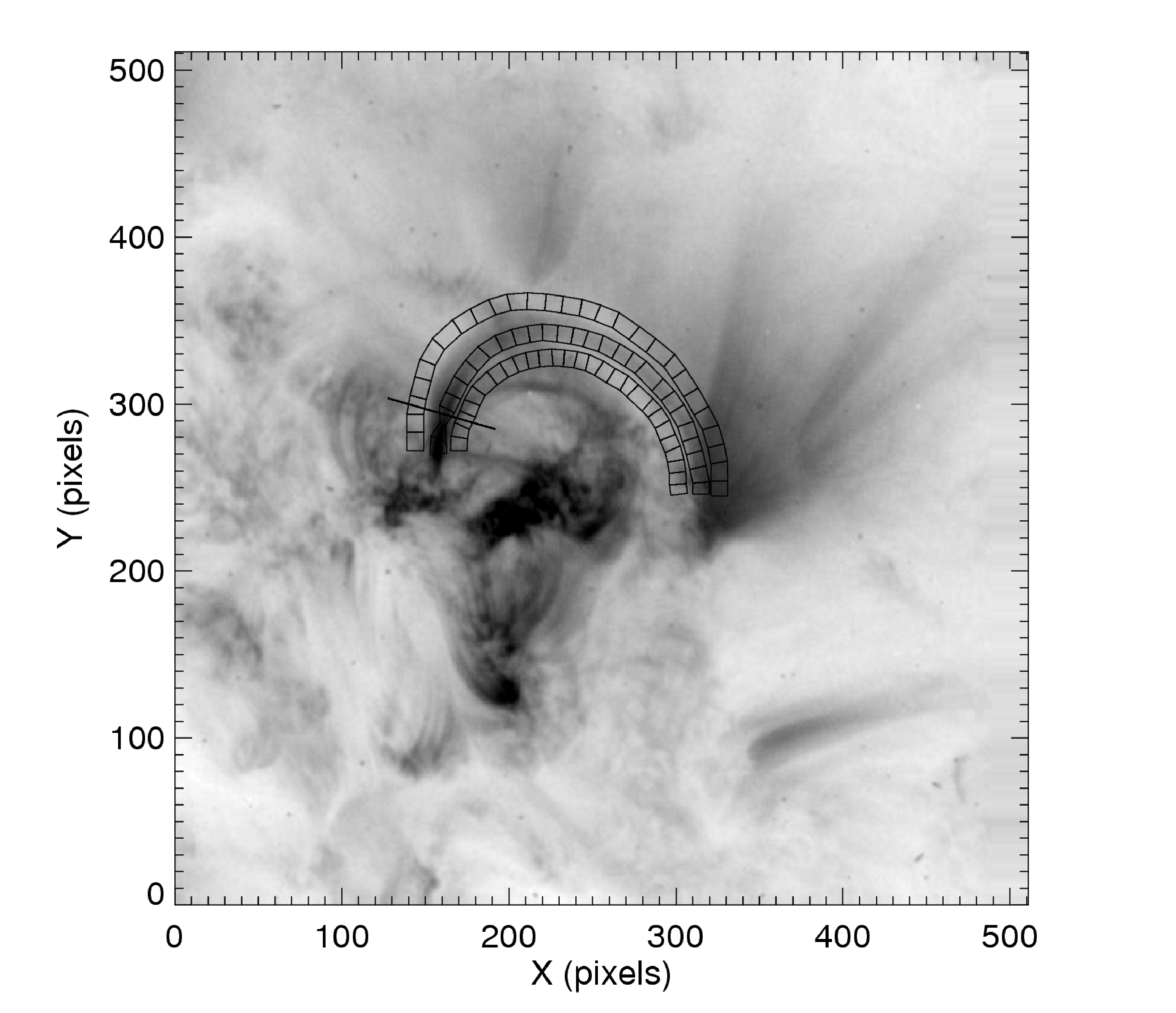}
	\includegraphics[width=9cm]{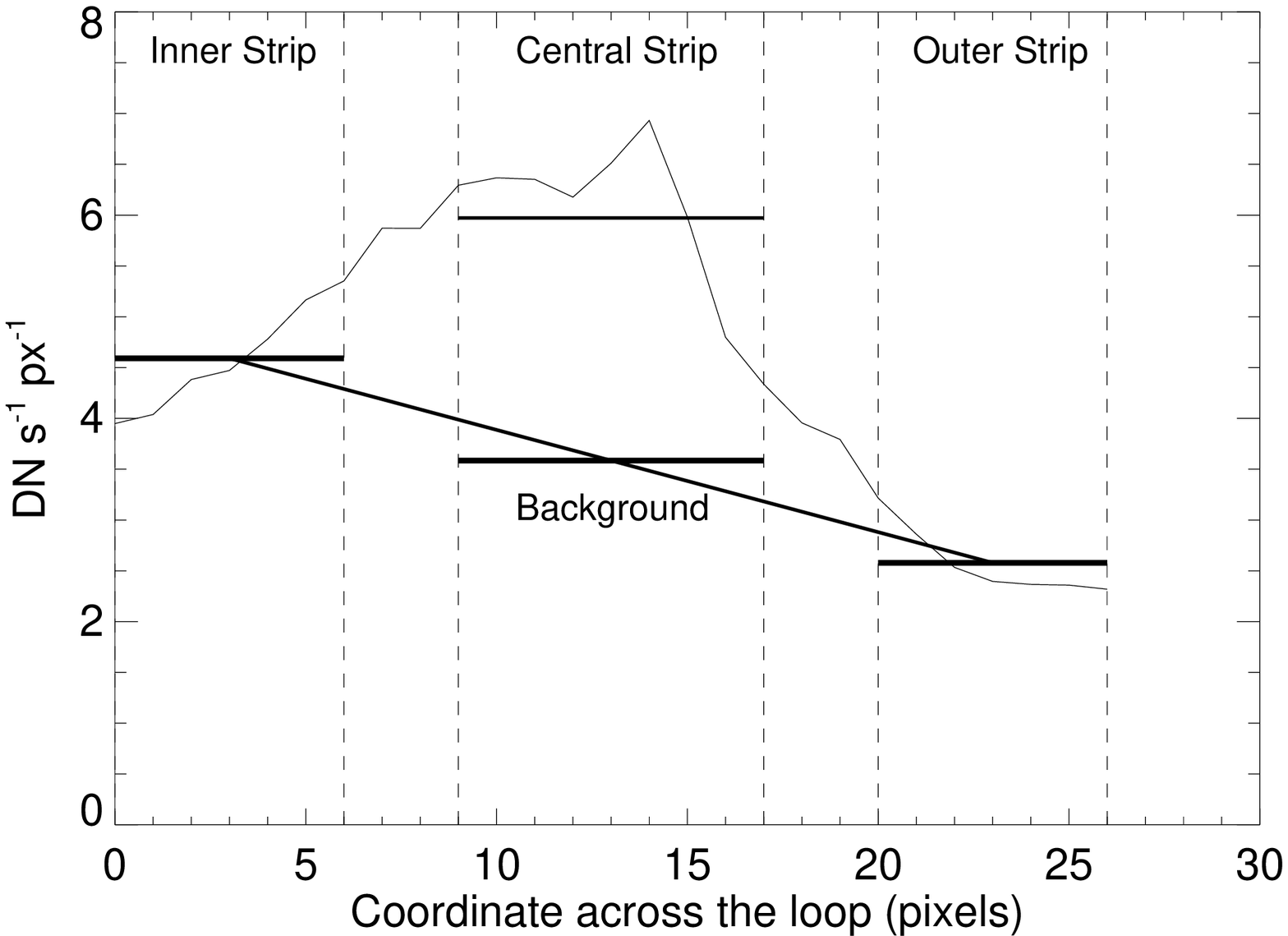}

\caption{ Background subtraction  with interpolation method. {\it Upper panel:} the strips for the loop analysis (central) and for the background subtraction (outer and inner) are marked in the loop region (07:39:26.000 UT, 171~{\AA}~ filter). {\it Lower panel:} Emission profile  ({\it thin solid line}) measured on a cut across the loop (marked in the upper panel). We compute the average emission (thick horizontal lines) in the corresponding sectors intersected by the outer and inner strips (bounded by vertical dashed lines). The background value (central horizontal line) is computed by linear interpolation of the average emission values of the outer strips. This is then subtracted from the total emission measured in the central strip (upper solid line).}
	\label{andamento_taglio}
\end{figure}

Here we compare the method already adopted in RC06 with an alternative and indepedent approach, based on interpolation between emission values in a region close to (but outside) the loop,  similar to those in \citealp{Testa_2002}, \citealp{Aschw_2005}, \citealp{Schmelz_2003} and \citealp{Aschw_2008ApJ}. As a first step, we measure the emission of the loop. We define a strip enclosing the loop in both TRACE filter passband, down to the visible footpoints and even beyond them, and divide it into sectors, as shown in Fig.\ref{andamento_taglio}. We have analyzed strips of different widths. A width of 10 pixels is a good compromise between too low statistics and overcoming too much the loop borders. With this choice we end up with 27 similar almost square sectors (RC06). Then, we mark two other strips, parallel and concentric to the one used to extract the loop emission, one outer and one inner of the imaginary circle which the loop is part of. We divide these two additional strips into sectors aligned to those of the central strip (Fig.\ref{andamento_taglio}). These two other strips are used to extract the emission for background subtraction. We put them as close as possible to the loop of interest, avoiding other structures near, although distinct from, the analyzed loop. Both strips also have a width of 10 pixels, and are divided into 27 sectors, so to have a one-to-one correspondence of these sectors with the sectors patching the loop. In each of the 27 sectors of the two external strips we compute the mean emission value per pixel. To each sector of the central strip we assign as background the emission value linearly interpolated between the two corresponding sectors of the external strips (Fig.\ref{andamento_taglio}).

For the sake of completeness, as a further refinement we have considered also a least-squares quadratic interpolation. We have then considered two further concentric strips around the analyzed loop, one outer and the other inner. However, the additional outer strips are so far from the central loop, that they end up to include other bright structures which severely contaminate the procedure and locally alter the otherwise smooth emission distribution. Therefore, in our case the higher order interpolation does not lead to an improvement of the results. Anyhow, we point out that our basic linear interpolation is made between values averaged over a sector, and therefore is already and intrinsically more accurate than a linear interpolation made between values in single pixels.

We remind that the method of background subtraction of RC06 is based on the fact that the loop disappears at the end of the image sequence. The last image (around 10 UT) is then subtracted pixel-by-pixel from all other images, assuming that the structures surrounding and crossing the loop of interest along the line of sight do not change much during the observation sequence. As an estimate of the fluctuations of the background during the observation, RC06 measured an average  pixel-by-pixel standard deviation of 13$\%$ both in the 171~{\AA}~ and in the 195~{\AA}~ filter band.

A more accurate subtraction would be obtained if the subtracted image resulted from the interpolation between an image before the loop switches on and one after the loop switches off. Unfortunately, 
this loop has a very long lifetime and in order to find images before the loop ignition we have to go back more than 25 ks earlier than the start time of our analysis. In our opinion, this time is too long to pretend that the environment did not change in the meantime and to have a reliable background, and we prefer to keep the final image as the only image for subtraction.

We also point out that the image to subtract was taken at 10:00 UT, significantly apart in time from the analyzed images taken between 06:30 and 08:30 UT. The perspective (i.e. viewing angle) of the observed loop was changing during that interval and this may impact the RC06 scheme which assumes that each line of sight intersects the same column of plasma throughout the entire sequence. However, we do not expect a significant effect since we estimate a change of the viewing angle of the order of $1-2^\circ$ or less, as confirmed by a visual inspection of the images after subtraction.

These are not the only two methods of background subtraction. \citealp{Schmelz_2007} and \citealp{Schmelz_2003} used two other lower-order methods of background subtraction: (1) constant background subtraction; (2) pixel pair background subtraction. In the former method they subtract a constant value extracted from a single background pixel chosen above the loop apex; in the latter they selected a background pixel for each loop pixel.

\section{Results}

\subsection{Loop Emission}

\begin{figure*}[]       %%%%%%%%%%%%%%%%%% FIGURE 5
	\centering	
	\includegraphics[width=7.5cm]{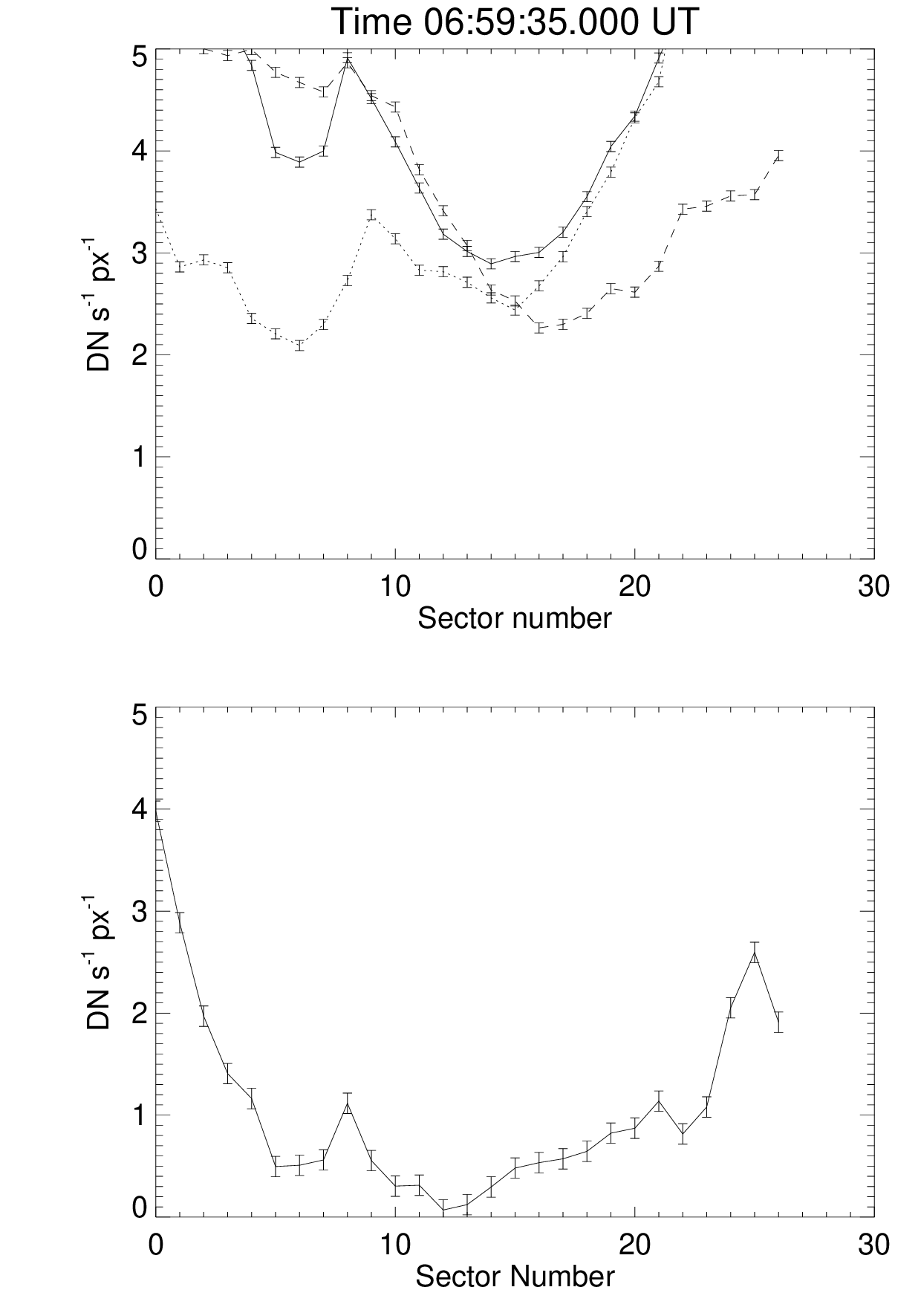}
	\includegraphics[width=7.5cm]{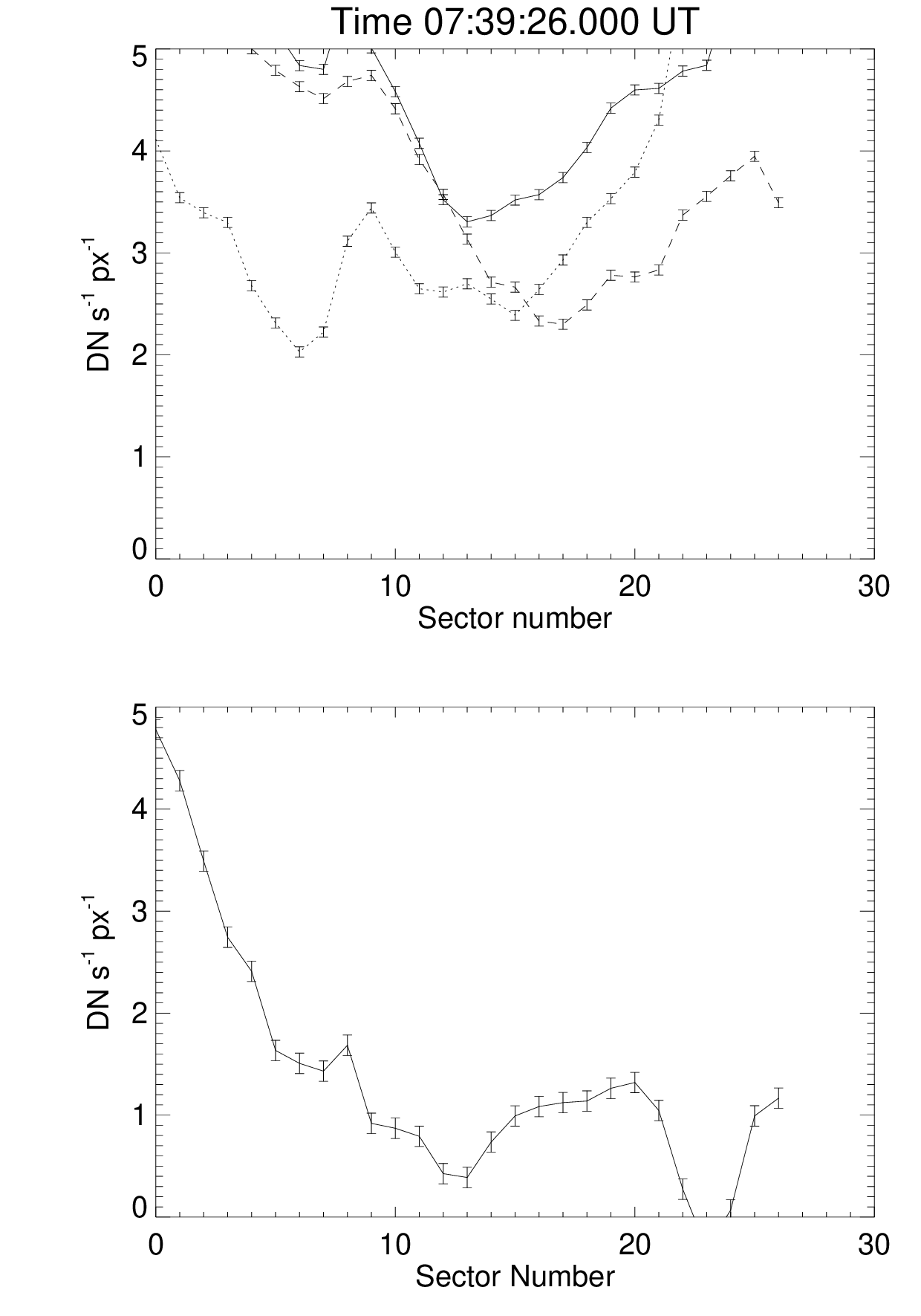}
\caption{Emission along the loop in the 171~{\AA}~ filter, at 06:59:35.000 UT (left column), and 07:39:26.000 UT (right column). Upper panels: emission along the loop strip (solid line), outer strip (dotted line), and inner strip (dashed line). Lower panels: emission after background subtraction with the interpolation method.}
	\label{anda_1}
\end{figure*}

\begin{figure*}[]       %%%%%%%%%%%%%%%%%% FIGURE 6
	\centering	
	\includegraphics[width=7.5cm]{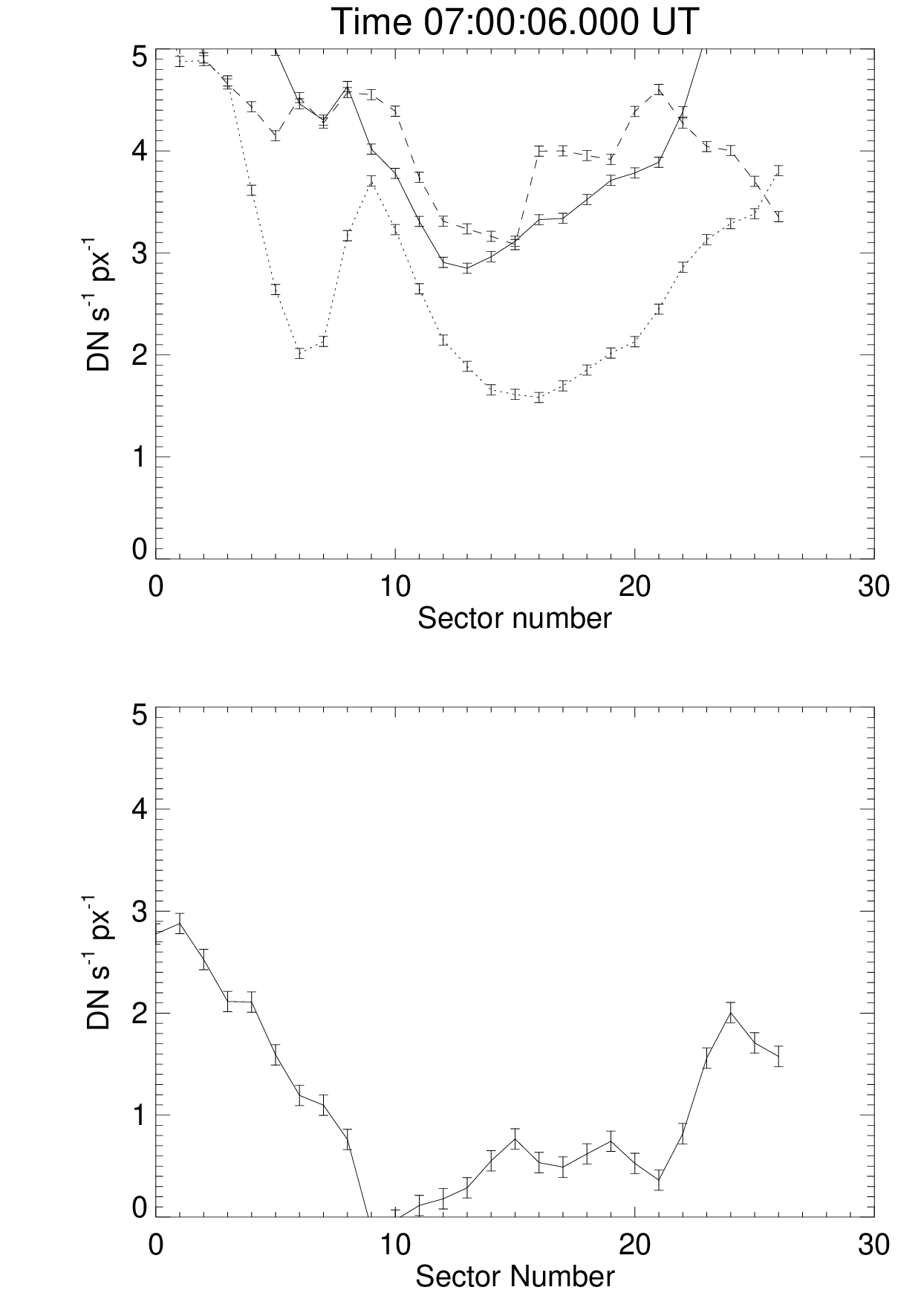}
	\includegraphics[width=7.5cm]{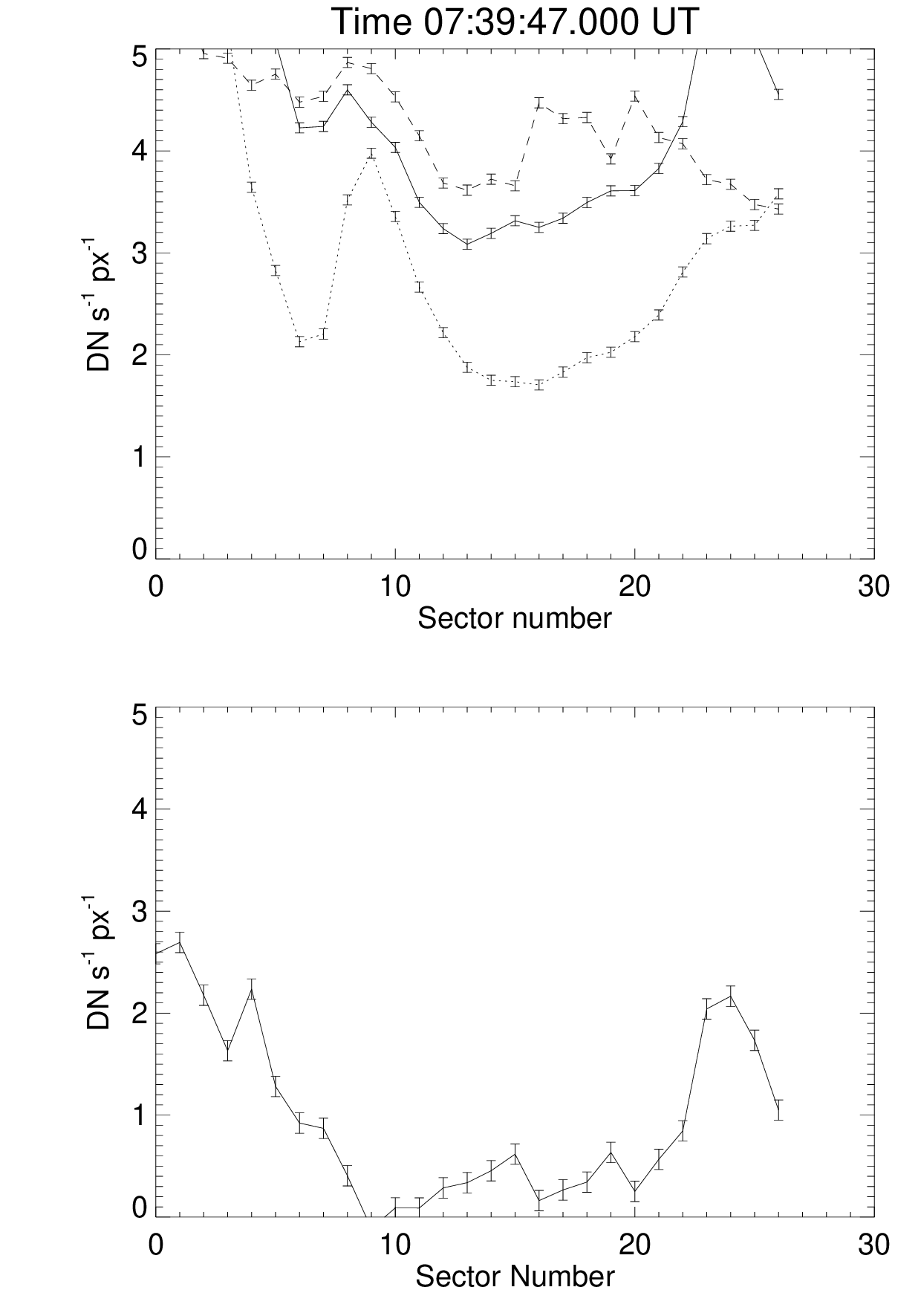}
\caption{As Fig.\ref{anda_1} for the 195~{\AA}~ filter, at 07:00:06.000 UT (left column), and 07:39:47.000 UT (right column).}
	\label{anda_3}
\end{figure*}

Fig.\ref{anda_1} and \ref{anda_3} show examples of the application of our interpolation method of background subtraction. The lines join the data points, obtained as mean values of the pixel emission in a sector.
The error of each mean value is estimated as the standard deviation of the mean in that sector. The error has been conservatively rounded to 0.05.
The lower panels of Fig.\ref{anda_1} and \ref{anda_3} show the results of the subtraction of values interpolating between the dotted and the dashed curves, from the solid curve.

\begin{figure*}[]       %%%%%%%%%%%%%%%%%% FIGURE 7
	\centering	
	\includegraphics[width=6.5cm]{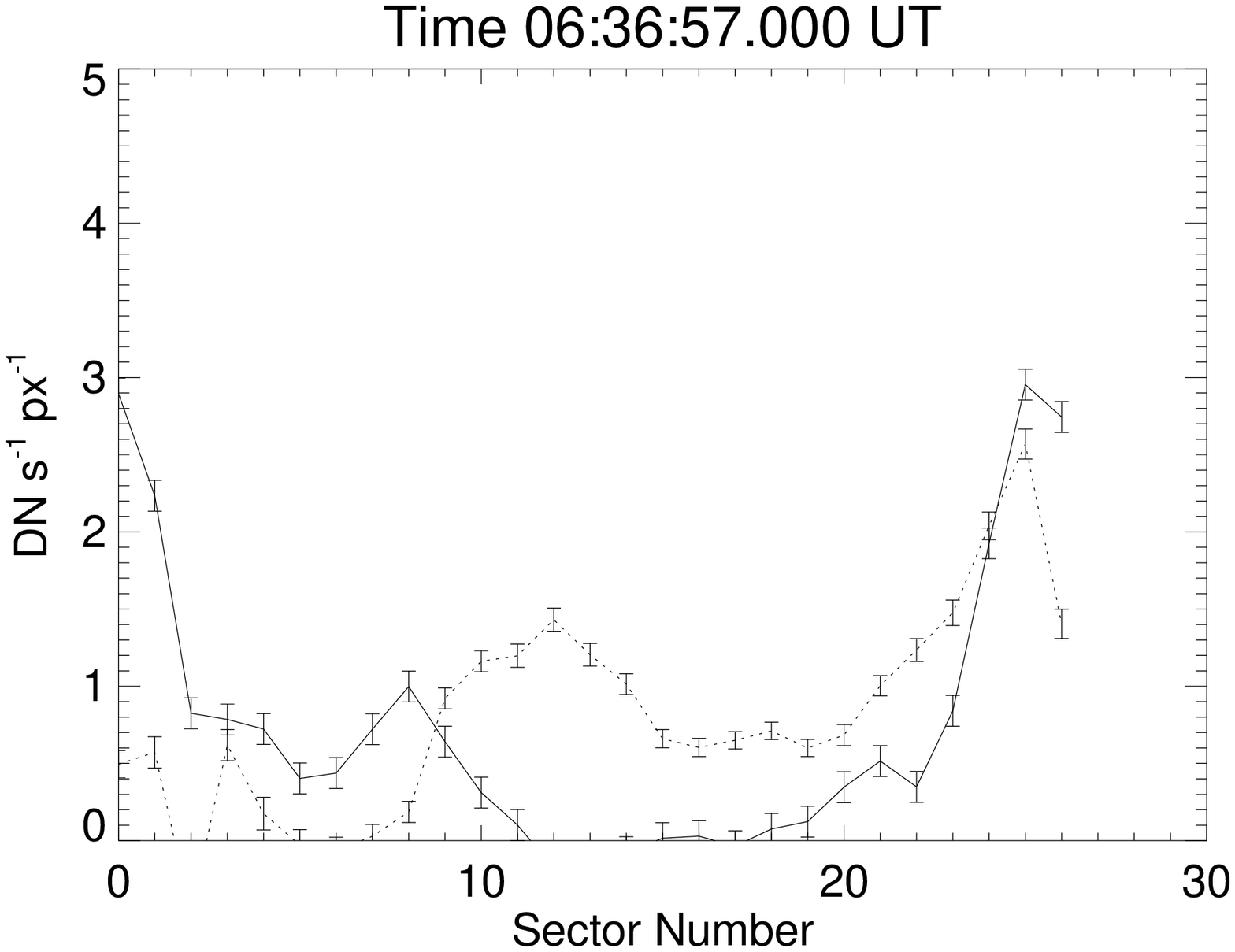}	\includegraphics[width=6.5cm]{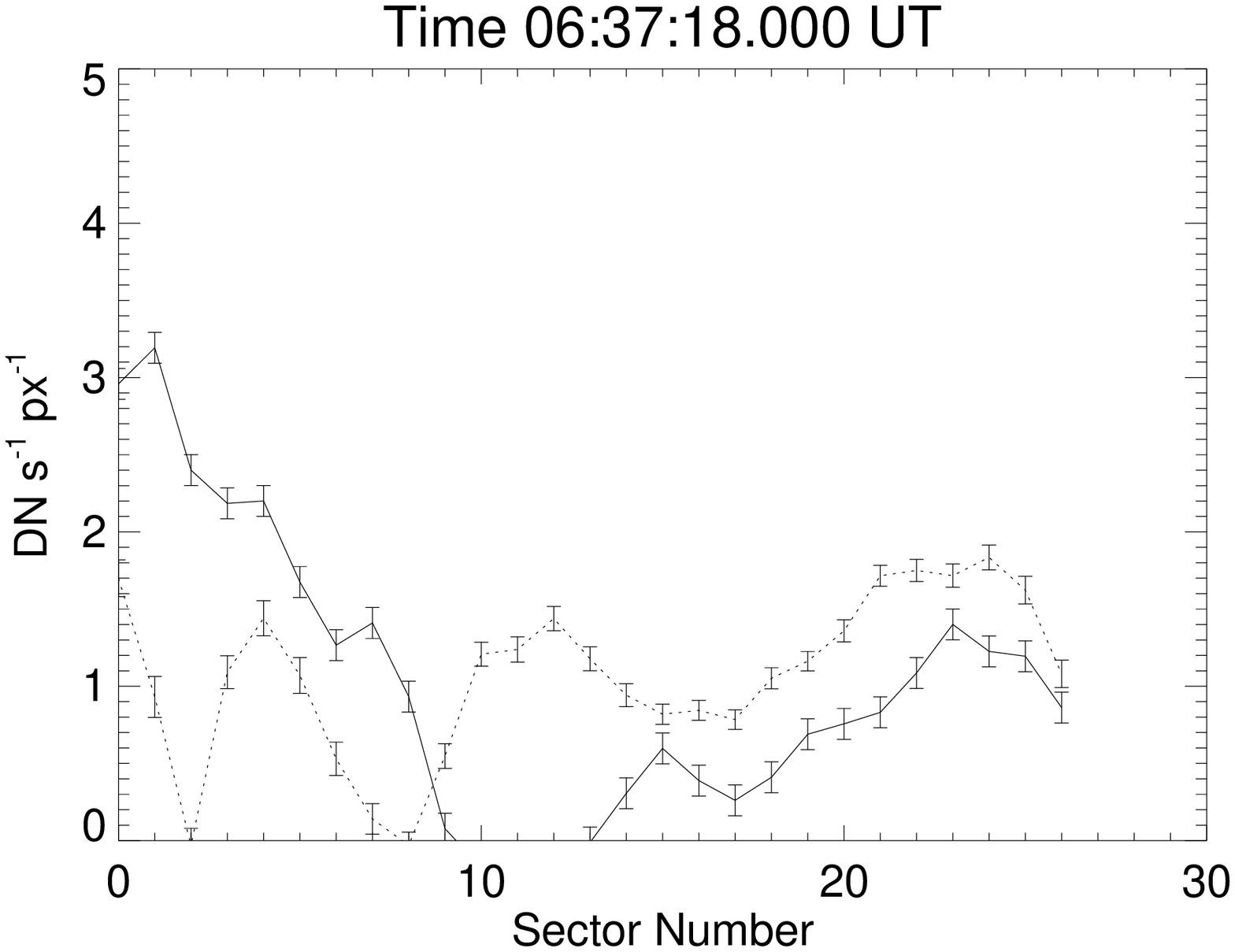}\\

	\includegraphics[width=6.5cm]{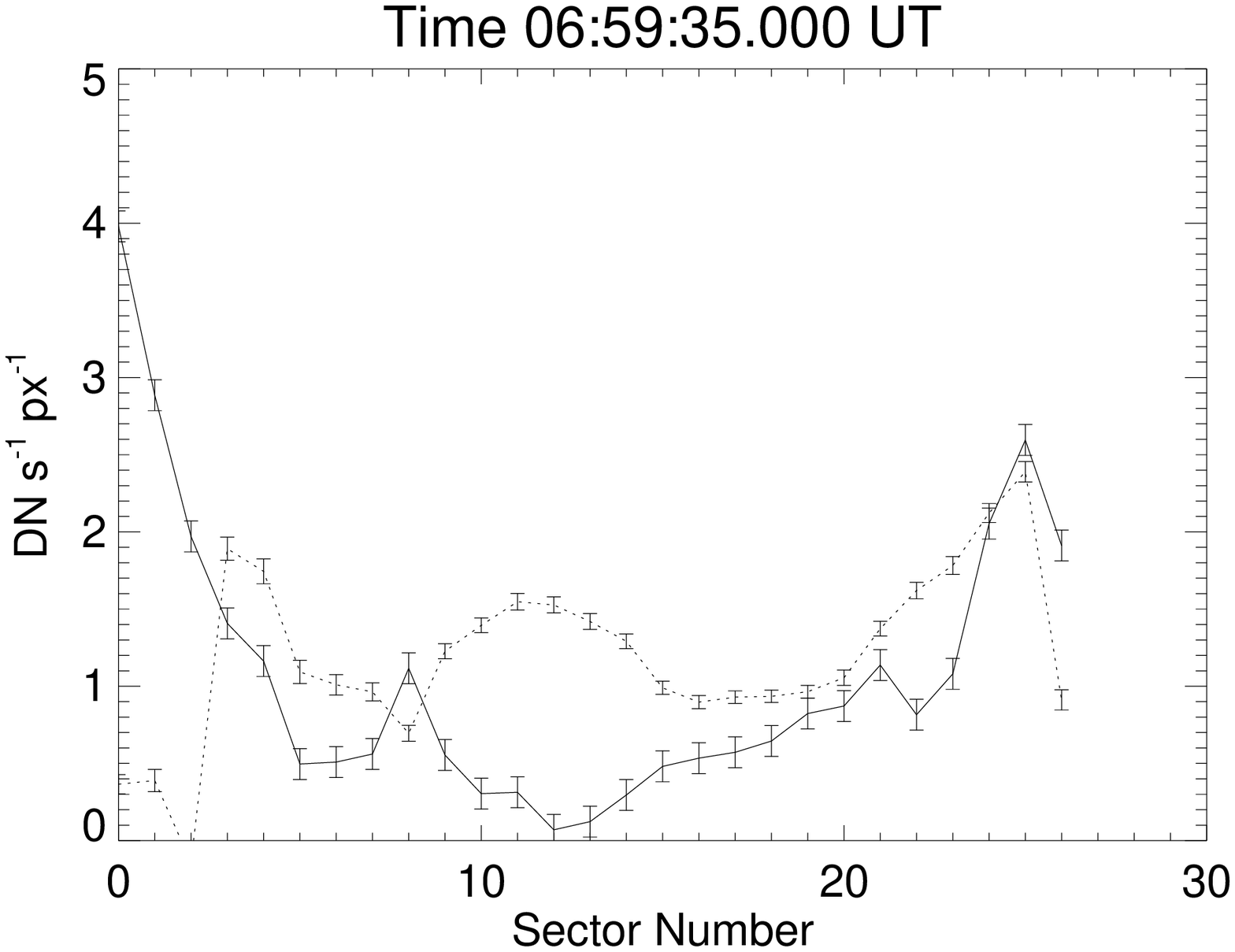}	\includegraphics[width=6.5cm]{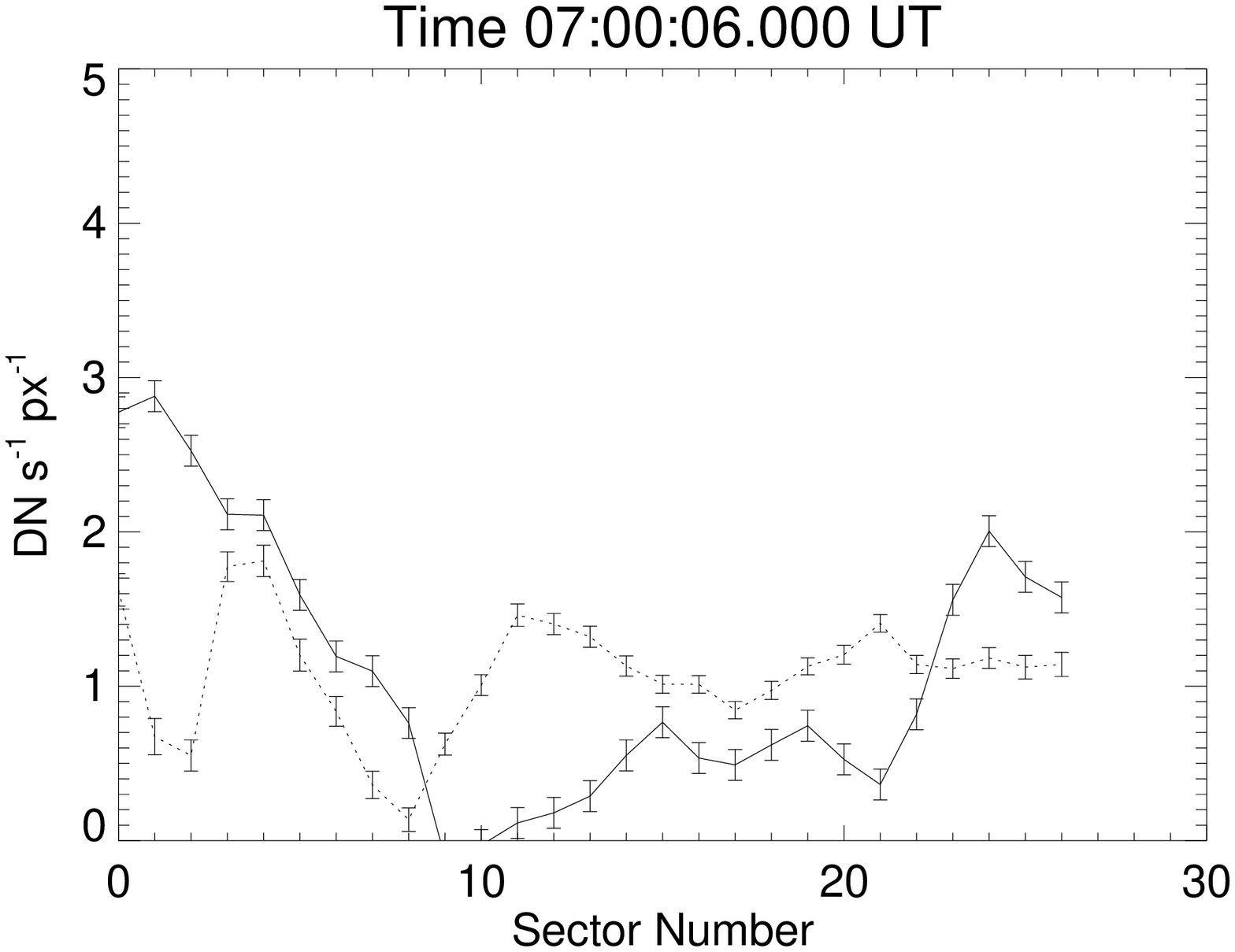}\\

	\includegraphics[width=6.5cm]{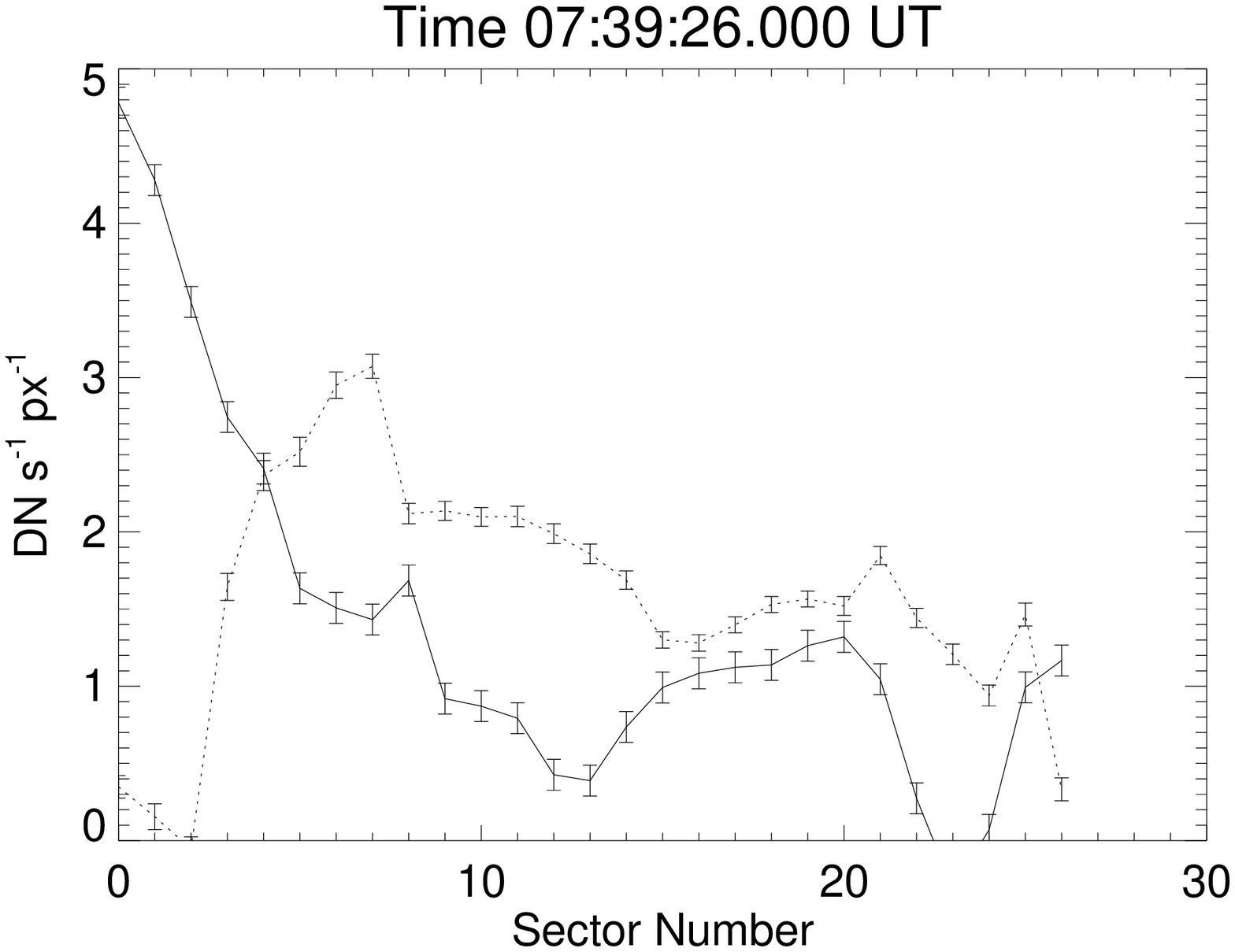}	\includegraphics[width=6.5cm]{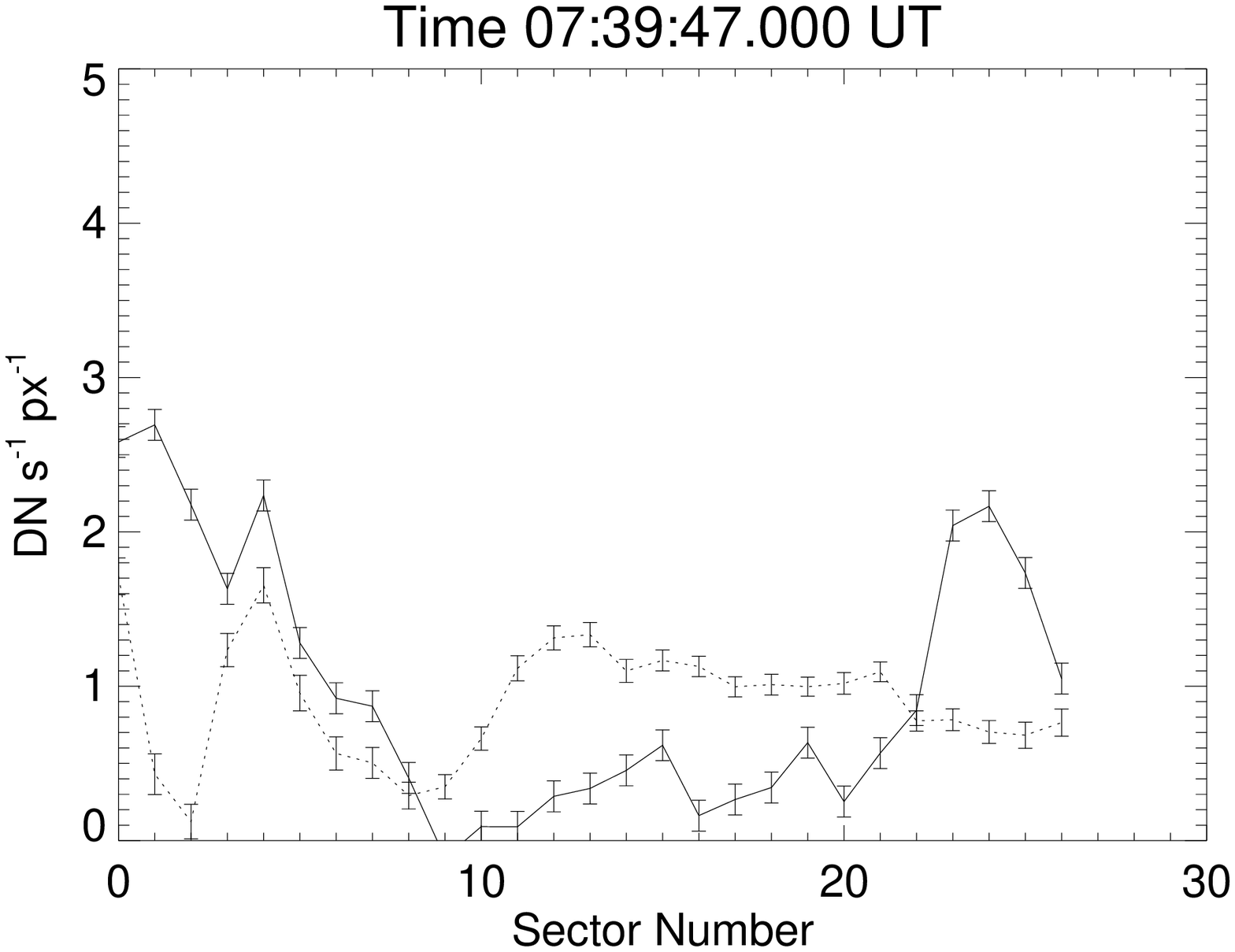}\\

	\includegraphics[width=6.5cm]{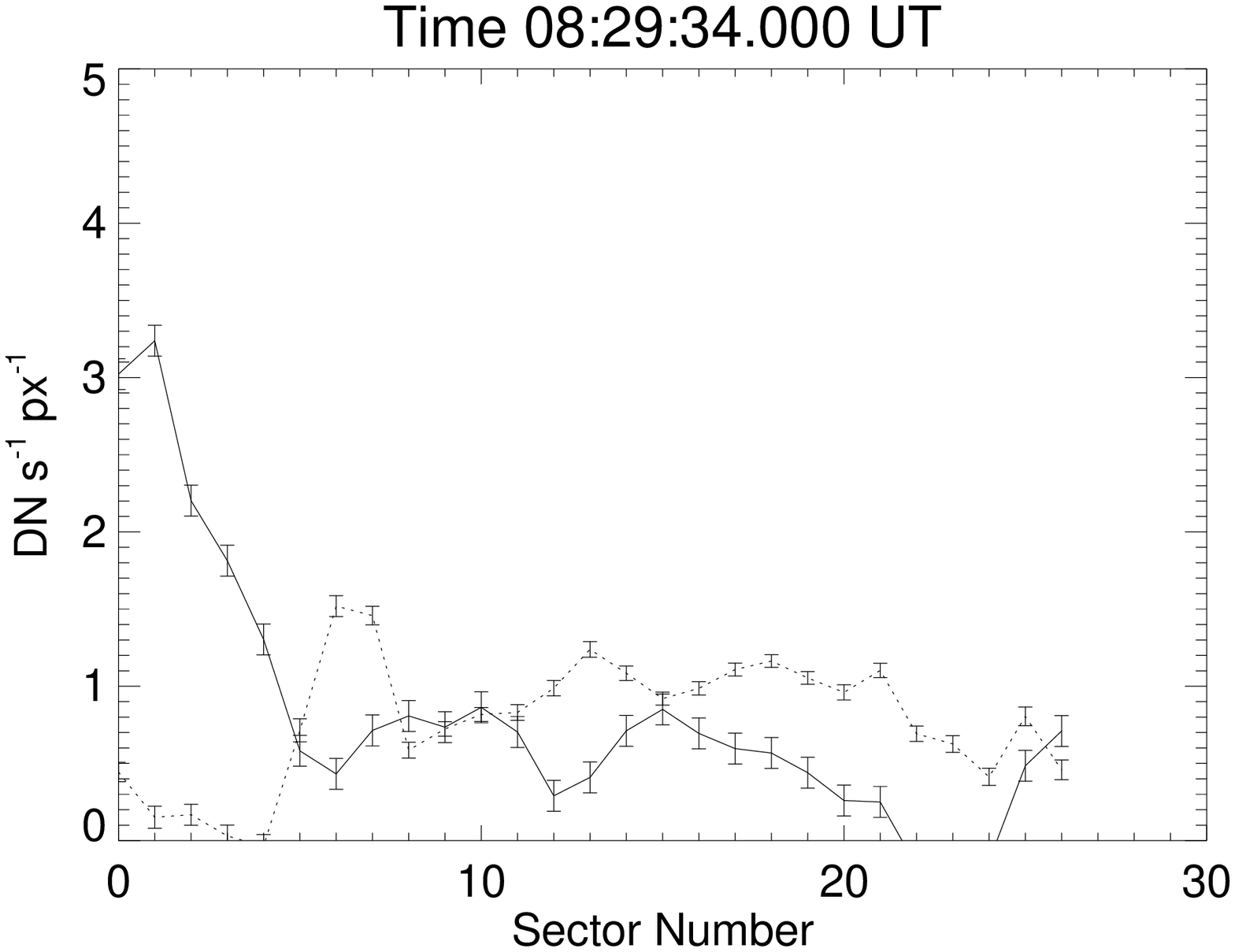}
	\includegraphics[width=6.5cm]{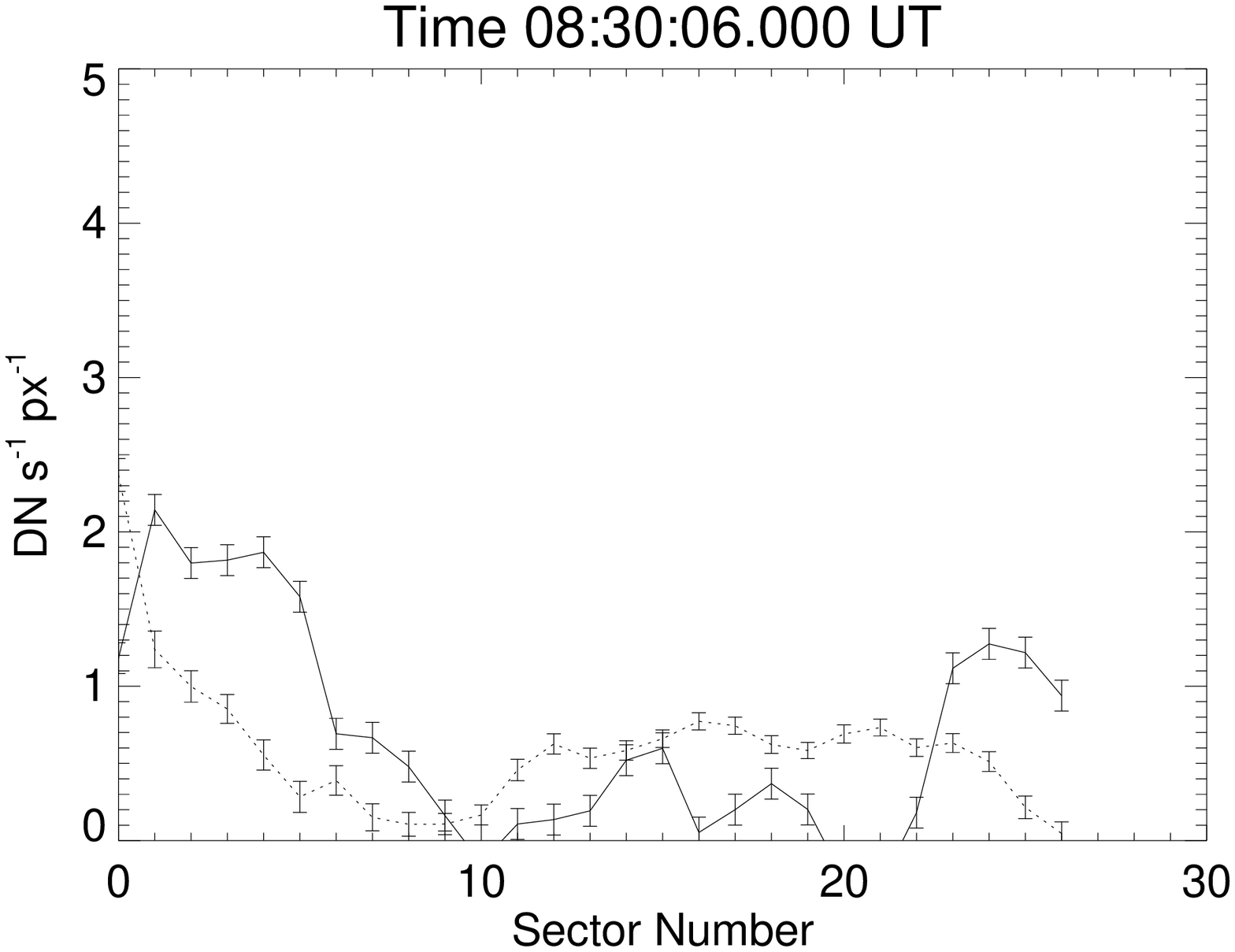}
	\caption{Emission along the loop at the labelled times in the 171~{\AA}~ filter (left column) and in the 195~{\AA}~ filter (right column). All panels show the emission along the loop, after background subtraction with two different methods: pixel-by-pixel method (RC06,dashed line), and interpolation method (solid line).}
	\label{confr_171}
\end{figure*}

The loop emission is significantly lowered by the background subtraction, in some points even completely cancelled, due to the overlap with other bright structure close to, but disantangled from our loop. For a direct comparison of this and the RC06 background subtraction methods, we now plot the emission along the loop obtained with the two methods in the same figure.  Fig.\ref{confr_171} shows the results in the 171~{\AA}~ filter and in the 195~{\AA}~ filter. We can see that the emission is significantly different after the background subtraction with the two methods: the profiles mostly differ in the central part, which is predicted to be very faint at any time after background subtraction by interpolation method. On the contrary, with the latter method the footpoints are found to be mostly brighter than with the pixel-to-pixel subtraction, although we see some agreement at some times (e.g. right end at 06:36 UT and 07:00 UT in both filters). While the result regarding the footpoints may be debated, since the footpoint regions cannot be clearly resolved in the original data (Fig.\ref{F-4panels-1}), there is no doubt that the bulk of the loop is clearly visible at many times (Fig.\ref{F-4panels-1} and Fig.\ref{ratio-panels}), and that. therefore, the emission cannot be so low, in contrast with interpolation method and in agreement with pixel-to-pixel method. Therefore, in the case where the two methods most differ, RC06's pixel-to-pixel method surely provides a more reliable result. There is some agreement between the profiles at the latest time.

Moreover, we notice in Fig.\ref{confr_171} that, since the subtracted emission obtained with interpolation method is invariably very low except near the loop footpoints, it does not show a clear evolution. Instead RC06 obtain that the background-subtracted central section of the loop has an evolution in agreement with the evolution observed in the images (Fig.\ref{F-4panels-1} and Fig.\ref{ratio-panels}). Again, therefore the results obtained with the method of RC06 appear more reliable.

\subsection{Temperature diagnostics}
\begin{figure}[]       %%%%%%%%%%%%%%%%%% FIGURE 8
	\centering	
	\includegraphics[width=9.cm]{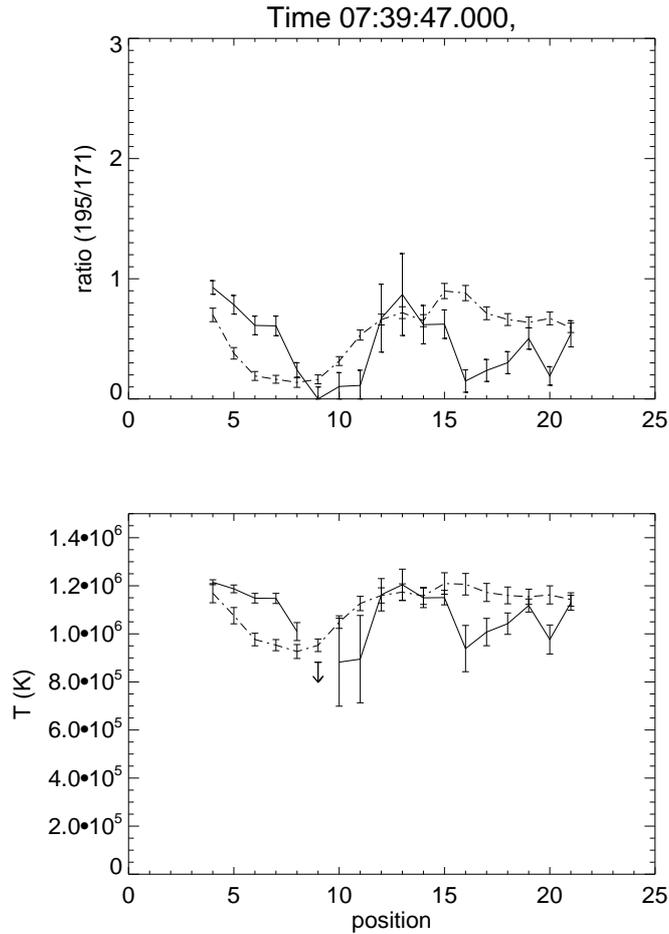}
	\caption{ Filter ratio 195/171 ({\it top}) and corresponding temperature ({\it bottom}) along the loop at 07:40 UT computed after the different background subtractions, pixel-to-pixel ({\it dashed}) and interpolation ({\it solid}). The arrow is an upper limit among interpolation data.}
	\label{fig:ratio_prof}
\end{figure}

To further assess the differences between the background subtraction methods, we have explored the temperature diagnostics obtained with filter ratio. It is well-known that for an optically thin plasma, isothermal along the line of sight, the ratio of the emission detected in two different filters is a function of the temperature only (e.g. \citealp{Rosner_1978ApJ}). From the ratio value measured in an image pixel we can then derive a temperature value in that pixel, and build therefore a proper thermal map. In the case of the narrow band 171~{\AA}~ and 195~{\AA}~ filters of TRACE, the relationship between temperature and filter ratio $195/171$ is monotonic only in a certain temperature range, i.e.  $0.7< T <1.8$ MK (e.g. \citealp{Aschw_2000ApJ}), which is also the range where the filters responses are the highest. So it is reasonable to assume that the detected plasma is all in that temperature range and this allows us to measure the temperature along the loop, after background subtraction. As an example, Fig.\ref{fig:ratio_prof} shows the filter ratio 195/171 and corresponding temperature along the loop at 07:40 UT computed after the  background subtraction with the pixel-to-pixel method (also shown in RC06) and with the interpolation method. It is immediately apparent that the ratio profile obtained with the latter method is much more irregular, so that we cannot define an overall trend. The other method instead leads to a more regular profile, coherent with the clear visibility of the loop in the filter ratio map (RC06), in spite of the dip on the left leg due to another crossing bright structure. We find similar trends in the corresponding temperature profiles along the loop. 

\begin{figure}[]       %%%%%%%%%%%%%%%%%% FIGURE 9
	\centering	
	\includegraphics[width=9.cm]{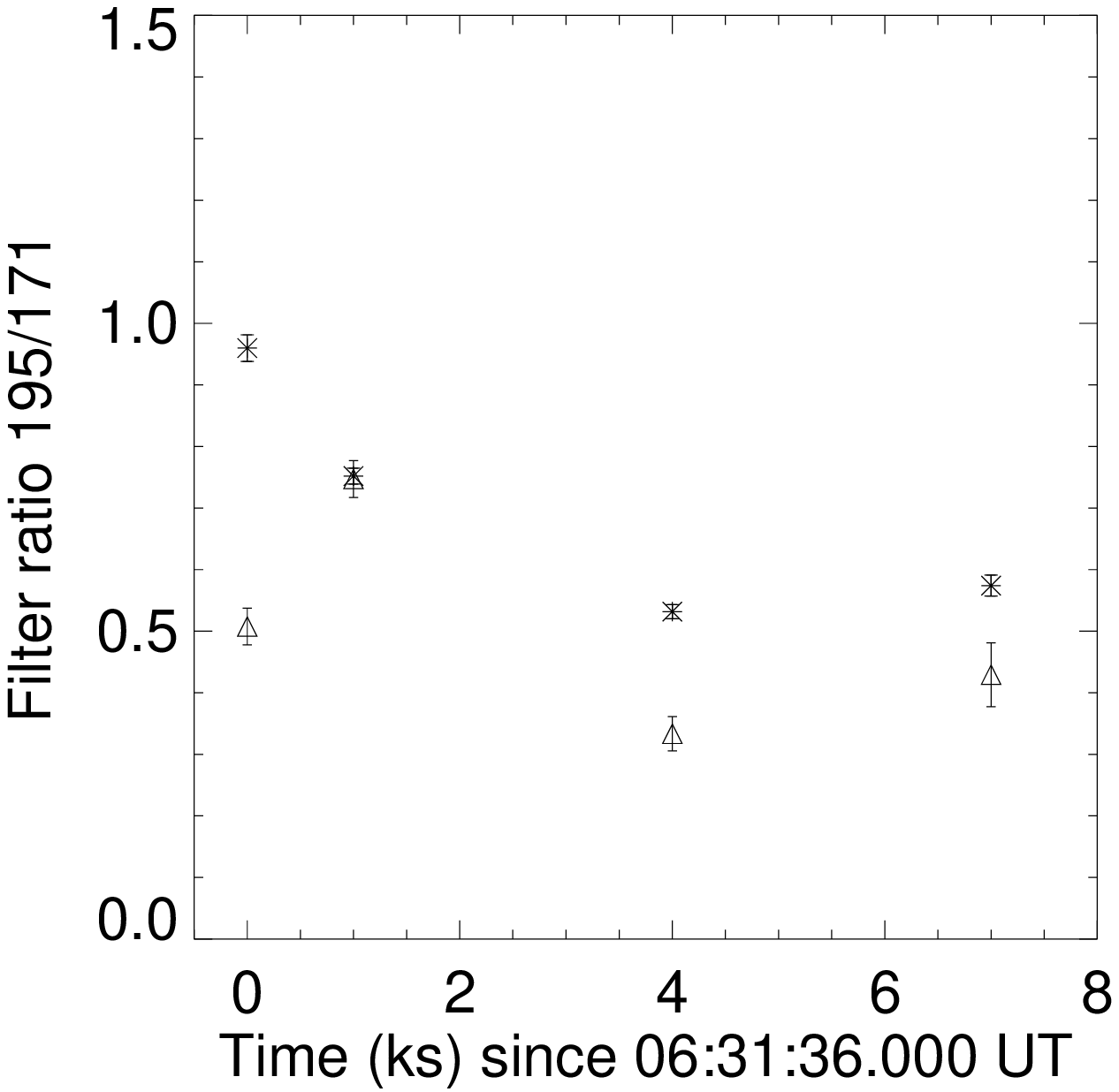}
	\caption{ Evolution of the loop average 195/171 filter ratio, sampled at the four times analyzed in this work and obtained after the different background subtractions, pixel-to-pixel ({\it stars}) and interpolation ({\it triangles}).}
	\label{fig:ratio_evol}
\end{figure}

We have to point out that, although both profiles indicate an overall small temperature variation along the loop, the pixel-to-pixel background makes the difference in identifying vs not identifying a trend. The globally flat temperature profile is a typical result from TRACE data (e.g. \citealp{Lenz_1999SoPh}) and has been explained partly with an instrumental bias (\citealp{Weber_2005ApJ}). However, RC06 pointed out that the filter ratio diagnostics can be anyhow meaningful and is sensitive enough to detect the loop progressive cooling. Fig.~\ref{fig:ratio_evol} shows the loop average filter ratio obtained after the two different background subtractions at four different times. From the evolution obtained with the interpolation method it is much less clear that the loop is cooling, confirming that the sensitivity of the temperature diagnostics is higher with the pixel-to-pixel background subtraction.

\section{Discussion}

The background subtraction is important for the analysis of TRACE coronal loops: the background signal is very high and influenced by the presence of many bright structures near and perhaps entangled with the analyzed loop, along the line of sight. Recent results (\citealp{DelZanna_2003}, \citealp{Testa_2002}, \citealp{Schmelz_2003}, \citealp{Aschw_2005}, \citealp{Reale_2006}, \citealp{Aschw_2008ApJ}) have established the importance of separating the actual loop plasma from the diffuse foreground and background emission that results from unresolved coronal structures and instrumental effects (stray light). Here we catch the unique opportunity to compare two different and independent methods of background subtraction: one that subtracts the interpolated emission of two off-loop strips; and the other that subtracts pixel-by-pixel the complete image after the disappearance of the loop (RC06). Extracting the background emission with two strips as near as possible to the loop of interest, we certainly consider and remove emission from structures that can intersect transversally the loop.
This is true as long as the ``contaminating''
loops run across the target loop, while it is not the case
if this happens at an oblique angle.

We have shown that not only do different methods lead to different emission profiles, but also the subsequent diagnostics is affected. With the pixel-by-pixel subtraction we are able to derive quite coherent filter ratio profiles along the loop and quite coherent temperature evolution: the loop is globally cooling as expected from the sequence of appearance/disappearance in the different TRACE filters. Such information is much less clear after using the other method of background subtraction.  

The RC06 background subtraction looks therefore a more reliable method. It is in principle very accurate, if the loop of interest is the most variable structure in the field of observation; it is direct, at variance from the interpolation method; it is applied pixel-by-pixel, allowing us to derive ``background-subtracted images'' and therefore to have a visual feedback, and to analyze all loop pixels, instead of sampling them at selected positions. On the other hand, this method works well as long as the structures surrounding and crossing the loop of interest along the line of sight do not change much during the observation.  In other words, the RC06 method cannot take time-variations of the background emission into account and we cannot exclude that crossing structures vary during the observation. We also remark that the pixel-by-pixel subtraction could be applied only because the loop disappears at the end of the image sequence. This condition can be matched only by evolving loops. For loops keeping steady during the entire observation, other methods must be used. The interpolation of out-loop emission has been often used (\citealp{Testa_2002}, \citealp{Aschw_2005}, \citealp{Schmelz_2003}, and \citealp{Aschw_2008ApJ}), but according to our analysis, it may lead to severe systematic errors because it may be affected by other structures close but distinct from the loop under analysis.

Fig.~\ref{andamento_taglio} points to the difficulties encountered by the interpolation
method since it shows that relatively close
to the target loop there is another structure with
similar intensity. Although the interpolation method should provide a good estimate of the background around the two loops, it is not able to take the overlap of the
two loops into full and proper account. A scheme which fits the two loops
separately together with a background estimated as above
may represent a better solution.

Other methods of background subtraction  are instead too ``operator-sensitive'', because they are based upon the meticulous selection of single background pixels, intentionally avoiding in this way structures that can cross the loop of interest and can alter significantly the loop emission. 

In conclusion, this work confirms and qualifies how a reliable background subtraction is a delicate and difficult task for the analysis of coronal loops observed with TRACE. The problem can be of course greatly reduced in observations with different instruments where the amount of instrumental background emission is lower and where therefore the presence of systematic effects has less influence.

\begin{acknowledgements}
We thank the anonymous referee for constructive suggestions. 
We acknowledge support from
Italian Ministero dell'Universit\`a e Ricerca and Agenzia Spaziale Italiana
(ASI), contract I/023/09/0. 
\end{acknowledgements}

% \bibliographystyle{aa} % style aa.bst
% \addcontentsline{toc}{section}{\bf Bibliografia}
% \bibliography{biblio} % your references Yourfile.bib
% \twocolumn 
% \addcontentsline{toc}{section}{\bf Bibliografia}

\bibliographystyle{aa} % style aa.bst
% \bibliography{biblio} % your references Yourfile.bib
\end{document}